\documentclass{aa}  

\usepackage{graphicx}
\usepackage{lipsum}
\usepackage{subcaption}
\usepackage{afterpage}
\usepackage{soul}
\usepackage{txfonts}
\usepackage[]{hyperref}
\newcommand{\ang}{\mbox{\normalfont\AA}}

\begin{document}

   \title{CN rings in full protoplanetary disks
   around young stars \\ as probes of disk structure}

   \subtitle{}

   \author{P. Cazzoletti \inst{1}
          \and
          E. F. van Dishoeck\inst{1,2}
          \and
          R. Visser\inst{3}
          \and
          S. Facchini\inst{1}
          \and
          S. Bruderer\inst{1}
          }

   \institute{Max-Planck-Institut f{\"u}r Extraterrestrische Physik, Gießenbachstraße, 85741 Garching bei München, Germany\\
              \email{pcazzoletti@mpe.mpg.de}
               \and
             Leiden Observatory, Leiden University, Niels Bohrweg 2, 2333 CA Leiden, The Netherlands
             \and
             European Southern Observatory, Karl-Schwarzschild-Strasse 2, 85741 Garching bei München, Germany
             }

   \date{xxxxx}

 
  \abstract
   {}
   {Bright ring-like structure emission of the CN molecule has been observed in protoplanetary disks. We investigate whether such structures are due to the morphology of the disk itself or if they are instead an intrinsic feature of CN emission. With the intention of using CN as a diagnostic, we also address to which physical and chemical parameters CN is most sensitive.}
   {A set of disk models were run for different stellar spectra, masses, and physical structures via the 2D thermochemical code DALI. An updated chemical network that accounts for the most relevant CN reactions was adopted.}
   {Ring-shaped emission is found to be a common feature of all adopted models; the highest abundance is found in the upper outer regions of the disk, and the column density peaks at 30-100 AU for T Tauri stars with standard accretion rates. Higher mass disks generally show brighter CN. Higher UV fields, such as those appropriate for T Tauri stars with high accretion rates or for Herbig Ae stars or for higher disk flaring, generally result in brighter and larger rings. These trends are due to the main formation paths of CN, which all start with vibrationally excited H$_2^*$ molecules, that are produced through far ultraviolet (FUV) pumping of $\rm H_2$. The model results compare well with observed disk-integrated CN fluxes and the observed location of the CN ring for the TW Hya disk.}
   {CN rings are produced naturally in protoplanetary disks and do not require a specific underlying disk structure such as a dust cavity or gap. The strong link between FUV flux and CN emission can provide critical information regarding the vertical structure of the disk and the distribution of dust grains which affects the UV penetration, and could help to break some degeneracies in the SED fitting. In contrast with C$_2$H or c-C$_3$H$_2$, the CN flux is not very sensitive to carbon and oxygen depletion.
}

\titlerunning{CN rings in full protoplsnetary disks}

   \maketitle
%

\section{Introduction}

Rotating disks of dust and gas around young stars are the cradles of
planets, but detailed studies of these objects have only been possible
in the last few years since the advent of the Atacama Large
Millimeter/submillimeter Array (ALMA). The resolution and sensitivity
of the data in the pre-ALMA era only allowed a handful of disks to be
characterized, most of the time only in dust \citep{williamscieza,2015PASP..127..961A}. The ALMA radio telescope  now opens up the possibility to survey
hundreds of disks and spatially resolve these disks in gas and dust
\citep{ansdell2016,2017AJ....153..240A,2016ApJ...827..142B, pascucci2016}.

One particularly intriguing feature of the new images are the ring-like
structures observed in dust
\citep{muto2012,andrews2016,isella2016} and gas
\citep{bruderer14,kastner2015,oberg2015,bergin2016}. In some cases,
such as the so-called transitional disks (TDs) with inner cavities, a
depletion of material in the central regions of the disk can lead to
ring-shaped structures for both solids and gaseous molecules
\citep[e.g.,][]{vandermarel2016}. However, it is apparent that gas and dust
can have different distributions and that molecular rings can be
due to peculiar physical and chemical conditions rather than to the morphological structure of the disk. In this sense, the morphological structure of the disk and
intensity of molecular line emission can provide a wealth of
information that cannot be obtained by dust emission alone.

In this scenario, CN is a particularly interesting molecule. One of
the brightest molecules after $^{12}$CO, with fluxes comparable to
those of $^{13}$CO, CN has been observed and studied in disks
both with single-dish observations
\citep[e.g. ][]{thi2004,kastner2008,salter2011,kastner2014}
and interferometric data \citep[e.g.,][]{dutrey1997,oberg2010,oberg2011,chapillon2012,guilloteau2014,teague2016,vt}. Although
none of the pre-ALMA era data had enough spatial resolution to give  information about the morphology of CN emission, CN has been used as a dynamical mass tracer \citep{guilloteau2014}.

\begin{figure}
\centering
  \includegraphics[width=.5\textwidth]{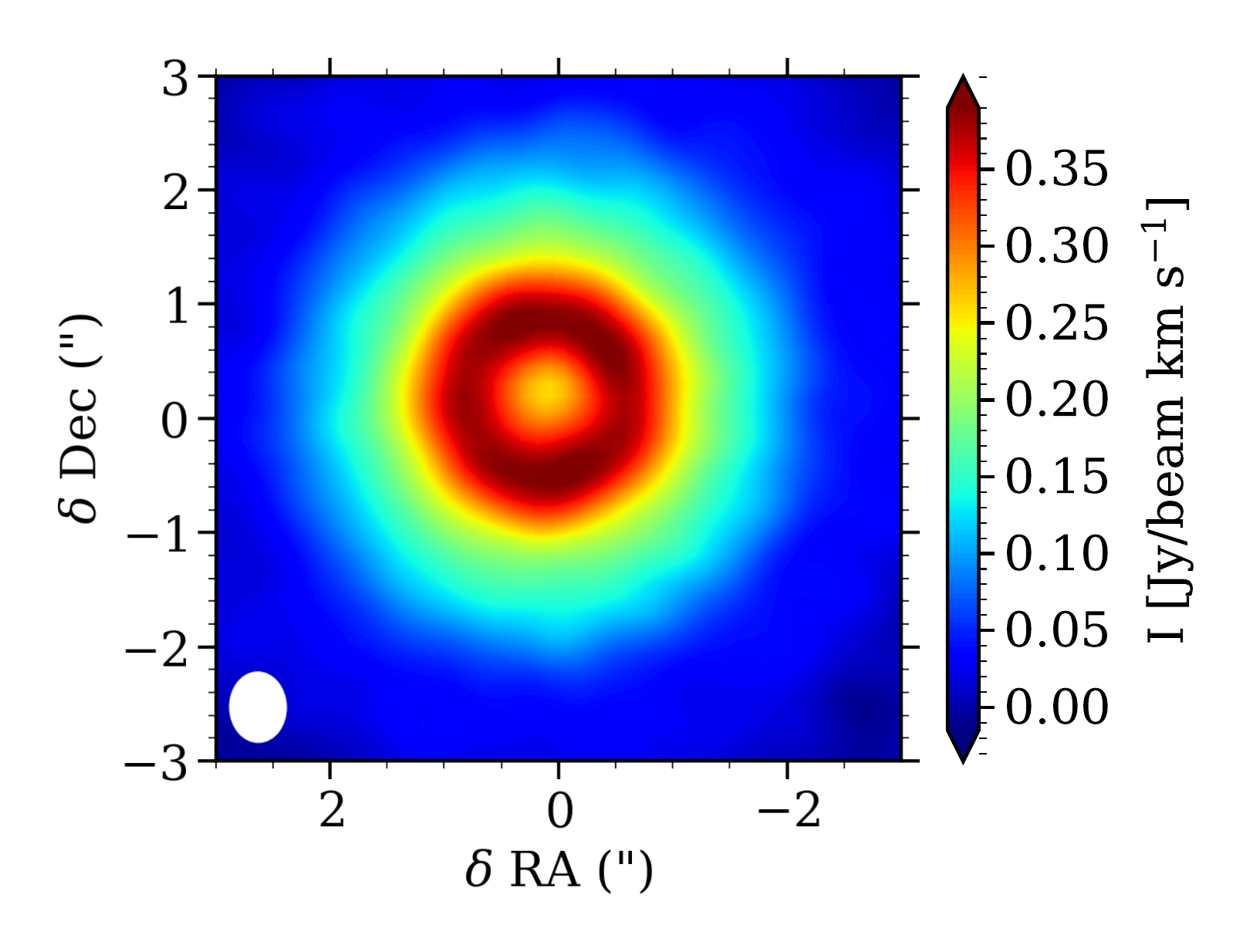}
\caption{CN ($N=2-1$) ring observed at 0.9" (47 AU)  in TW Hya with ALMA \citep[data from][]{teague2016}. The white ellipse at the bottom-left corner shows the synthesized beam size. }\label{fig:twhya_teague}
\end{figure}

CN emission is expected to be sensitive to UV radiation, since
its formation relies on the existence of atomic C and N or can result
from photodissociation of HCN. Modelling of CN in photodissociation
regions (PDRs) \citep[e.g. ][]{jansen1995,sternberg1995,2010ApJ...722.1607W} and disks
\citep[e.g. ][]{vz2003} has been performed and confirms this
expectation, placing CN in the upper warm molecular layers. For these
reasons, CN is also referred to as a PDR tracer and could thus be a sensitive probe of disk structure. 

With ALMA, more information about this molecule is rapidly
gathered. The CN ($N=2-1$) and CN ($N=3-2$) transitions are observable
with Band 6 and Band 7 respectively, and the emission can now be
spatially resolved. Observations at both wavelengths have provided new
information, in particular about the morphology. \citet{teague2016} used
CN ($N=$2-1) hyperfine transitions data to measure the
velocity dispersion in the TW Hya disk. At the same time, the
beam size of $0.5''\times0.42''$ allowed \citet{teague2016} to clearly resolve and
distinguish a ring-like structure in their data (Fig. \ref{fig:twhya_teague}). CN was also detected in many disks the ALMA Lupus disk survey \citep{ansdell2016}. Two of these disks, namely \mbox{Sz 98} and \mbox{Sz 71}, were large and bright enough to show resolved ring structures,  as presented in \citet{vt}.

Ring-shaped emission has also been observed for other PDR tracers,
such as C$_2$H. \citet{bergin2016} concluded that in order to have
ring-shaped C$_2$H emission in full (i.e., non-transitional) disks, a
[C]/[O]>1 ratio is required, which at the same time also
explains the high observed fluxes \citep{kama2016}.  In this work, we investigate whether
ring-shaped emission is an intrinsic feature of CN in all
protoplanetary disks or whether it is due to the specific underlying
morphology of some particular disks, such as transitional disks. By modelling CN chemistry we also want to gain further insight into
what kind of physical and chemical information about the disk structure can be inferred from resolved CN
observations, i.e. to which parameters the CN flux and ring size
are most sensitive.

\begin{table}[tbh]
\caption{Parameters of the disk models in the main grid.}
\label{tab:parameters}
\centering
\begin{tabular}{ll}
\hline\hline
Parameter	 & Range \\
\hline
\emph{Chemistry} \\
$\rm [C]/[H]$ & 1$\times 10^{-4}$\\
$\rm [O]/[H]$ & 3.5$\times 10^{-4}$\\
$\rm [N]/[H]$ & 1.6$\times 10^{-5}$\\
$\rm [PAH]$ & $10^{-3}$ ISM abundance \\
\\
\emph{Physical structure}& \\
$\gamma$ & 1\\
$\psi$ & 0.1, 0.2, 0.3 \\
$h_{\rm c}$ & 0.1, 0.2  rad \\
$R_{\rm c}$ & 60 AU \\
$M_{\rm gas}$ & $10^{-5} , 10^{-4} , 10^{-3}, 10^{-2}, 10^{-1} M_\odot$ \\
$f_{\rm large}$ & 0.99\\
$\chi$ & 0.2 \\
\\
\emph{Stellar spectrum}&\\
$T_{\rm eff}$ & 4000 K +  UV ($\dot M=10^{-8}\,\rm M_\odot/year$),\\
&10000 K\\
$L_{\rm bol}$ & 1, 10 $L_{\odot}$\\
$L_{\rm X}$ & $\rm 10^{30}\, erg\,s^{-1}$\\
\\
\emph{Dust properties}&\\
Dust & 0.005-1 $\mu$m (small)\\
& 1-1000 $\mu$m (large)\\
\emph{Other parameters}&\\
Cosmic-ray ionization& \\
rate per H$_2$ & $5\times10^{-17}{\rm s}^{-1}$ \\
 External UV flux & $G_0$\\
\hline
\end{tabular}
\end{table}

Section \ref{sec:model} describes the code, adopted model, physical
structure, and chemical network. In Section
\ref{sec:results} the results of our models are presented: the main
observed trends and the most important reactions responsible for the
ring-shaped CN emission. Finally, in Section \ref{sec:discussion} our
results are compared with the data available to date, and the main
information that can be obtained from CN is highlighted.

\section{Model}\label{sec:model}

\begin{figure*}[h!]
\begin{subfigure}{0.51\textwidth}
\centering
\includegraphics[page=1,width=1\linewidth]{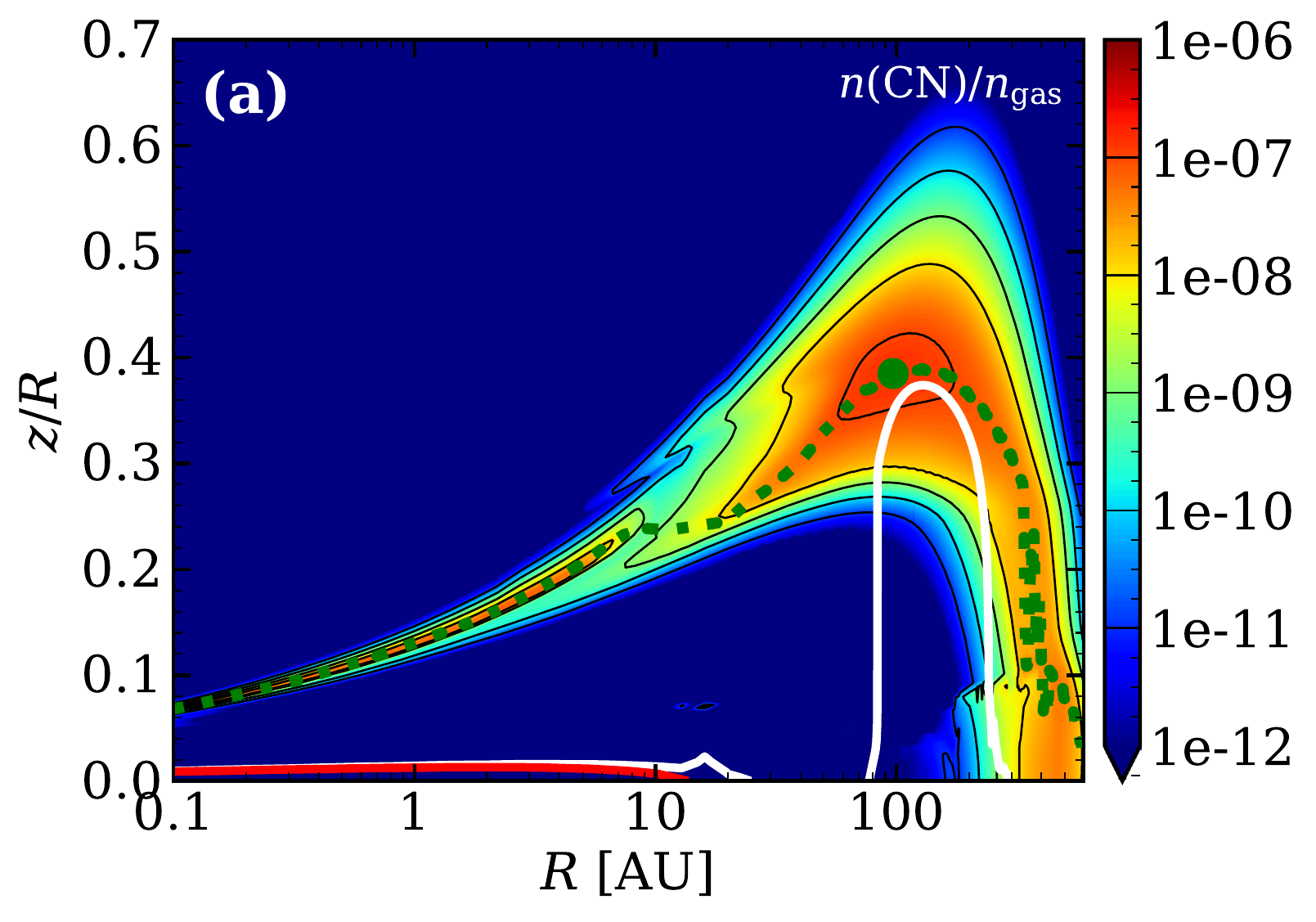}
\phantomcaption\label{fig:cnabu} 
\end{subfigure}
\begin{subfigure}{0.49\textwidth}
\centering
  \includegraphics[width=1\textwidth]{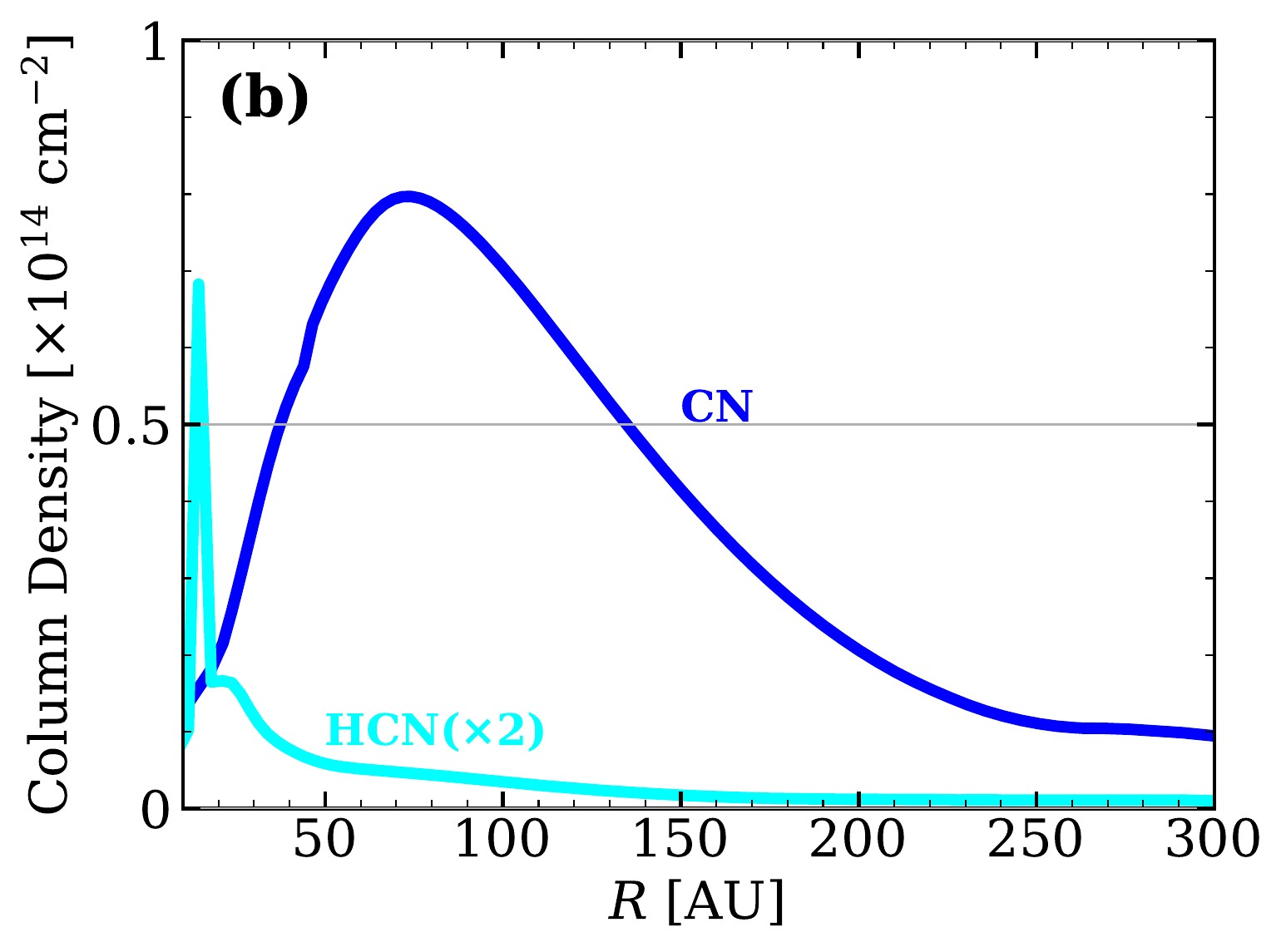}
\phantomcaption\label{fig:cncolumndens}
\end{subfigure}
\caption{(a) Abundance of CN in a $10^{-2}\,\rm M_{\odot}$, $\psi=0.3$, $h_{\rm c}=0.1$ disk surrounding a T Tauri star. The location of the peak value is indicated with the green dot. The shape of the distribution of the outer high-CN-abundance region is representative for all the other models. The white line shows the $ \tau=0.5$ contour of the CN ($N=3-2$) emission line. The red line shows the dust $\tau=1$ layer at the same wavelength.  The green dotted line indicates the location of the vertical CN abundance maximum for each radius. (b) CN (solid blue line) and HCN (cyan line) column densities calculated for the same model, plotted on a linear rather than logarithmic scale. The HCN column density was multiplied by 2 and is more radially concentrated than CN.}
\end{figure*}

In this work, we modelled the disks and CN emission using the 2D thermo-chemical code DALI, which combines radiative transfer, chemistry, thermal balance and ray-tracing calculations.  This code has been described in detail and extensively tested with benchmark problems  \citep{bruderer12,bruderer13} and its results have been compared to observations in a series of other papers \citep{bruderer12,fedele13,bruderer14,miotello2016,2017arXiv170506235F}. Given an input gas and dust density structure, the dust radiative transfer problem is solved via a 2D Monte Carlo method to obtain the dust temperature structure $T_{\rm dust}$  and the mean continuum intensities at wavelengths ranging from the UV to mm. Using an initial guess for the gas temperature, the abundances of the molecular and main atomic species are then calculated with either a time-dependent or steady-state chemical network simulation, and their excitation is computed through non-LTE calculations. Finally, the gas temperature structure $T_{\rm gas}$ can be obtained from the balance between the cooling and heating processes: the new $T_{\rm gas}$ is then used to re-calculate the chemistry, and the last two steps are iterated until a self-consistent solution is obtained. Both continuum emission and spectral image cubes are finally created using a ray tracer.

\subsection{Adopted physical model}
For our disk models we assumed a simple parametrized surface density distribution as in \citet{andrews2011}, following an exponentially tapered power law:
\begin{equation}\label{eq:dens}
\Sigma_{\rm gas}(R)=\Sigma_{\rm c}\bigg(\frac{R}{R_{\rm c}}\bigg)^{-\gamma}\exp \Bigg[-\bigg(\frac{R}{R_{\rm c}}\bigg)^{2-\gamma}\Bigg],
\end{equation}
where $R_{\rm c}$ is a characteristic radial scale for the surface density profile and $\Sigma_{\rm c}$ sets the normalization of the density profile. 
This surface density profile corresponds to the so-called self-similar solution to the viscous accretion disk model \citep{1974MNRAS.168..603L,1998ApJ...495..385H}, where the shear viscosity parameter is assumed to depend on the radius via $\nu\propto R^{-\gamma}$ and not to change in time.

The vertical gas density distribution follows a Gaussian, with an aspect ratio  that depends on the radius as 
\begin{equation}
h(R)=h_{\rm c}\bigg(\frac{R}{R_{\rm c}}\bigg)^{\psi}. 
\end{equation}
In order to mimic dust settling, two populations of grains are considered following the approach of \citet{2006ApJ...638..314D}: a small-grains population ($0.005-1\,\mu\rm m$) and a large-grains population ($1-1000\,\mu\rm m$). The small-grains population has a scale height of $h$, while the scale height of the more settled large-grains population is reduced to $\chi h$, where $0<\chi<1$. The fraction of dust surface density distributed in the two populations is controlled by the $f_{\rm large}$ parameter via the relations $\Sigma_{\rm dust,large}=f_{\rm large}\Sigma_{\rm dust}$ and \mbox{$\Sigma_{\rm dust,small}=(1-f_{\rm large})\Sigma_{\rm dust}$}. We assumed that most of the dust mass in concentrated into the more settled dust grains \citep[e.g.,][]{testi2003,rodmann2006,lommen2009,ricci2010}. After some preliminary testing, we concluded that varying the $f_{\rm large}$ settling parameter between 0.9 and 0.99 does not produce any significant difference in CN emission. Consequently, we kept $f_{\rm large}$ fixed in our grid and assumed \mbox{$f_{\rm large}=0.99$} in all of our models. Also, negligible variation was observed in both the abundance and emission profiles when the settling parameter $\chi$ is varied between 0.1 and 0.4. In this work, $\chi=0.2$ was used.

Other important parameters that affect the chemistry in the models are the stellar and interstellar radiation fields, and the cosmic-ray ionization rate. For the chemistry of CN, the FUV part of the spectrum ($6-13.6 \,\rm eV$) becomes particularly important, since photons in this energy range can be responsible for both the formation of CN (through HCN photodissociation) and for its destruction \citep[e.g., ][]{jansen1995,sternberg1995,visser2016,heays2017}. We modelled different stellar spectra to study the effect of different UV fluxes by assuming the stars to emit as a black body at a given effective temperature $T_{\rm eff}$ with an excess UV due to accretion.The stellar X-ray spectrum was modelled as a black body at $7\times 10^7\,\rm K$ between 1 and 100 keV. The interstellar radiation field and the cosmic microwave background were also accounted for and considered to be isotropic. We set the cosmic-ray ionization rate per H$_2$ to $5\times10^{-17}\,\rm s^{-1}$.

\begin{figure}
  \centering
  \includegraphics[page=4,width=\linewidth]{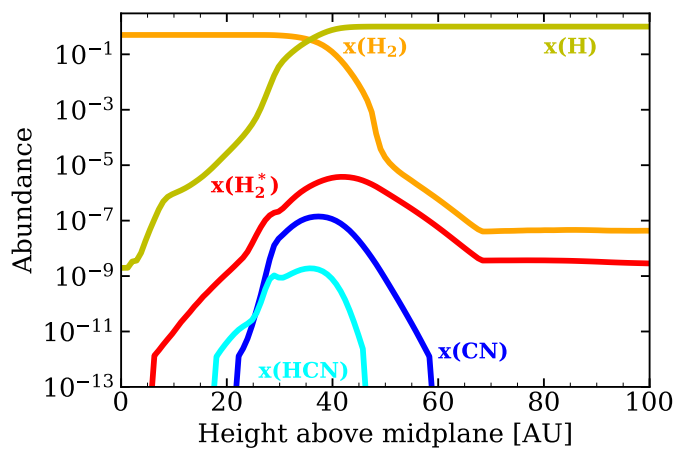}
  \caption{Vertical cut at the radial location of the CN peak abundance. The CN peak is located between the molecular hydrogen (H$_2$) region and the atomic layer.}\label{fig:Vertical_cuts}
  \end{figure}

\begin{figure*}
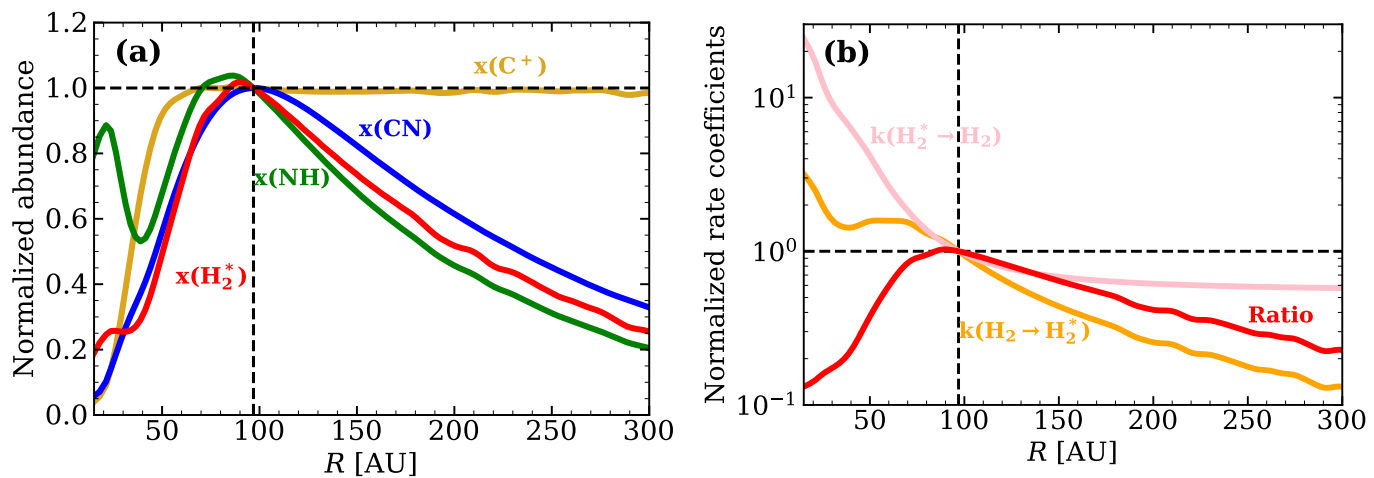

\centering
\begin{subfigure}{0.49\textwidth}
  \centering
  \includegraphics[page=2,width=\linewidth]{./images/figure_group_1.pdf}
  \phantomcaption\label{fig:radial_abus}
\end{subfigure}
\begin{subfigure}{.49\textwidth}
  \centering
  \includegraphics[page=3,width=\linewidth]{./images/figure_group_1.pdf}
  \phantomcaption\label{fig:h2_cd_exc}
\end{subfigure}
\caption{(a) Radial abundances for the most important reactants in the formation of CN (see reactions \ref{eq:nhform} and \ref{eq:cnplusform}), along the dotted green line in Fig. \ref{fig:cnabu}. Each curve is normalized to 1 at the location of the CN maximum. (b) H$_2$ excitation rate, H$_2^*$  de-excitation rate and their ratio along the  dotted green line in Fig. \ref{fig:cnabu}.  All quantities are normalized to their value at the CN peak abundance (green dot in Fig. \ref{fig:cnabu}) and refer to the model shown in Fig. \ref{fig:cnabu}. The dashed vertical line indicates the location of the CN peak abundance.}
\end{figure*}

The computation was carried out on a spatial grid of 265 cells in the radial direction and 120 in the vertical direction. In the radial direction, 35 cells were logarithmically spaced in the first $20\,\rm AU$ and 230 are linearly spaced at $\sim2.5\,\rm AU$ resolution in the outer disk. The cells in the vertical direction were linearly spaced and their size scales with the aspect ratio $h$ at a given radius. A spectral grid was also used for the radiative transfer, consisting of 58 wavelength bins extending from $912\,\ang$ to $3\,\rm mm$.  The abundance for CN and the other chemical species were then derived by assuming steady-state chemistry for the chemical calculations. Such an approach is justified by the fact that the chemistry in the upper layers is fast, therefore making the relevant timescales much shorter than the average age of a Class II disk.

\subsection{Grid of models}

The goal of our work is to investigate whether or not the ring-like CN emission observed in spatially resolved disks is a common feature in all protoplanetary disks, or if it is due to the specific global morphology of some particular disks such as TDs. The origin of such rings is also investigated. In addition, it is also clear observationally that CN emission can be very bright and it is important to understand which disk and stellar properties most strongly affect the CN line emission intensity. We ran a small grid of models varying some of the parameters describing the physical structure of the disk, dust distribution, and the stellar spectrum. A summary of the parameters used in the models can be found in Table \ref{tab:parameters}.

\subsection{Chemical network}

Our models use the chemical network from \citet{bruderer12}, expanded
and updated by \citet{visser2016} to produce accurate abundances for
HCN, HNC and CN\@. In particular, this network includes all reactions
from the cyanide chemistry review by \citet{loison2014} and the updated CN photodissociation rate from \citet{heays2017}. Isotopes are
not considered.

The network contains standard gas-phase reactions, photoionization and
photodissociation, reactions induced by X-rays and cosmic rays. some
reactions with PAHs, and a limited grain-surface chemistry. Full
details can be found in Appendix A.3.1 from \citet{bruderer12} and
\citet{visser2016}. Gas-grain interactions include freeze-out,
sublimation, and photodesorption. The binding energy of CN is set to
1600 K, as recommended by the Kinetic Database for Astrochemistry
\citep[KIDA;][]{wakelam12}.

In the surface layers and outer disk, the UV radiation field is strong
enough to pump H$_2$ into vibrationally excited states, denoted as
H$_2^*$. This vibrational energy can enable neutral-neutral reactions
that would otherwise have insurmountable activation barriers
\citep{london78,tielens1985,sternberg1995,bruderer12}. Of particular
importance to our models is the reaction of atomic N with H$_2^*$ to
form NH, which in turn reacts with C or C$^+$ to form CN or CN$^+$
(see below). The initial step has an activation energy of 12\,650 K
\citep{davidson90a} and would normally only proceed at gas
temperatures above $\sim$1200 K\@. However, the UV pumping of H$_2$ in
the surface layers allows the N + H$_2^*$ reaction to proceed
regardless of temperature. As such, it is a crucial first step in the
formation of CN\@.  The H$_2$ pumping rate is  taken to be eight times its photodissociation rate \citep{sternberg2014}.

The chemistry is run in steady-state mode, assuming
initial molecular abundances as in \citet{cleeves2015} but with
[C]/[H] increased to $10^{-4}$ from $10^{-6}$ unless specified
otherwise. Because the surface layer dominates the CN abundance, its
chemistry and emission are not found to be sensitive to the specific
period over which the chemistry is run and results are similar in
steady state.

\afterpage{

\begin{figure*}[h!]
\centering
\begin{subfigure}{0.49\textwidth}
  \centering
  \includegraphics[width=\linewidth]{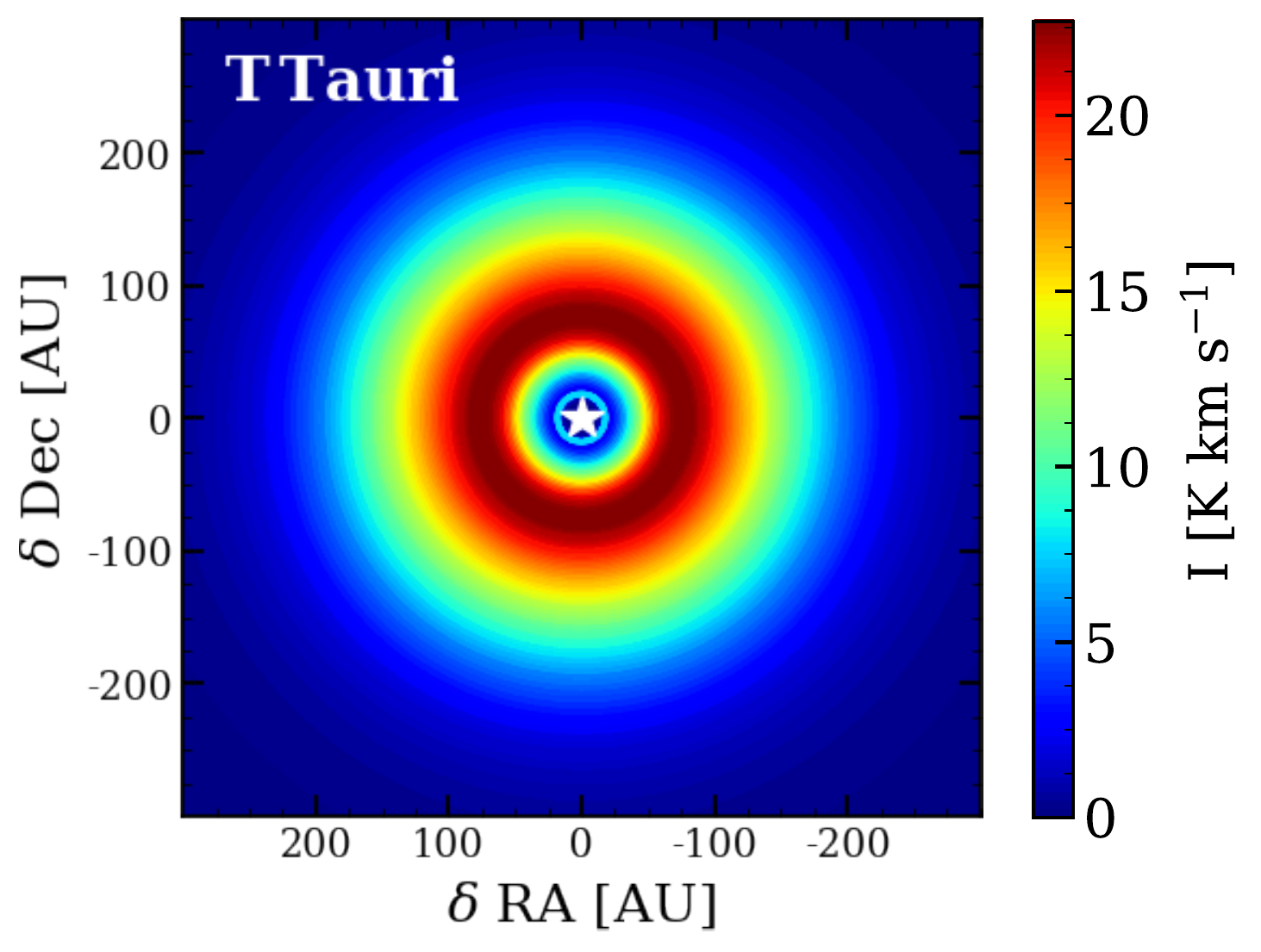}
  \phantomcaption\label{fig:ring_image_ttau}
\end{subfigure}
\begin{subfigure}{.49\textwidth}
  \centering
  \includegraphics[width=\linewidth]{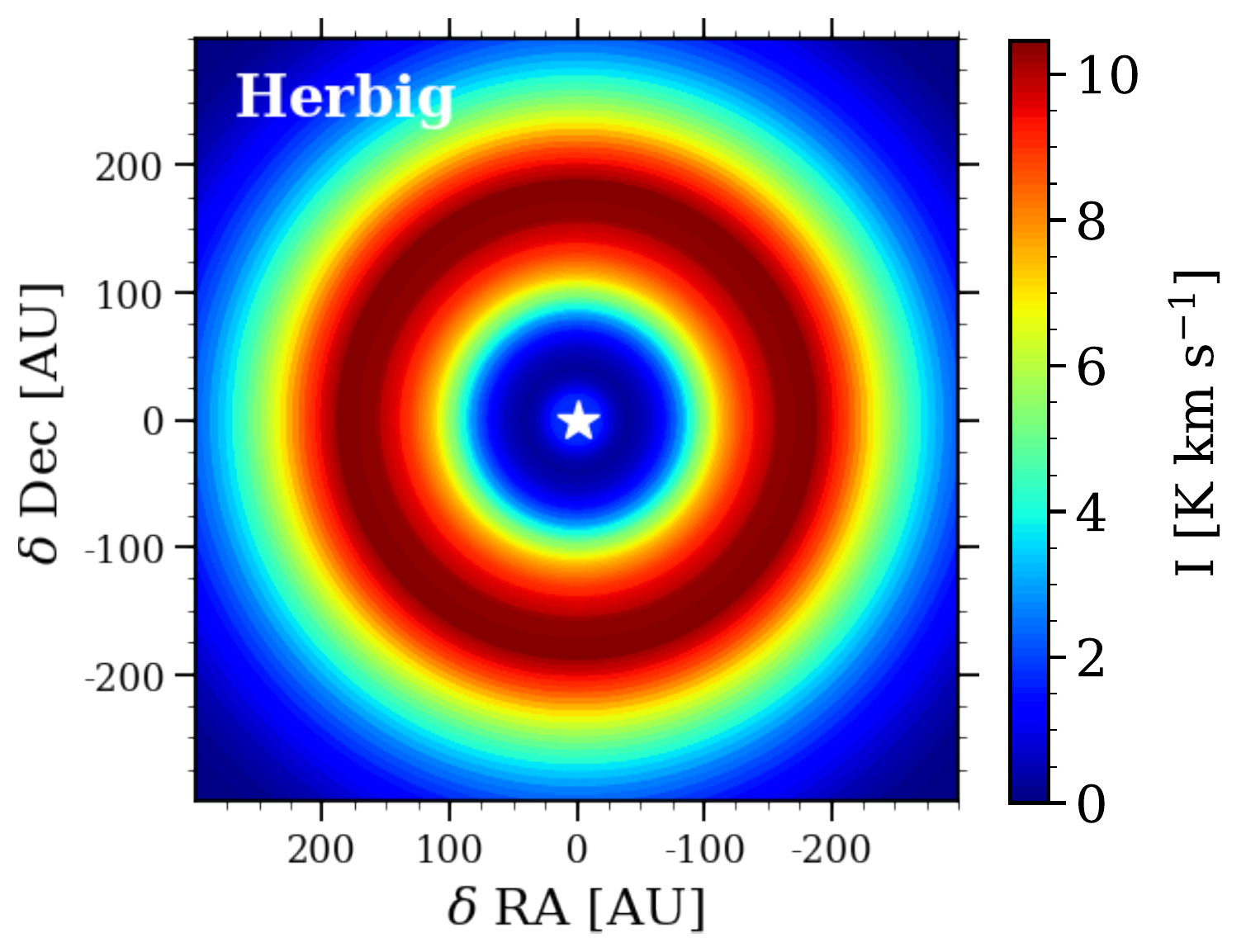}
  \phantomcaption\label{fig:ring_image_herbig}
\end{subfigure}
\caption{Ray-traced images of the CN emission rings in models with $10^{-2}\,\rm M_{\odot}$, $\psi=0.3$, $h_{\rm c}=0.1$ disks around a T Tauri and a Herbig star. The ring in the Herbig disk is clearly more extended than that around the T Tauri star. The images are not convolved with any beam, and the resolution is set by the size of the cells in the spatial gridding of the disk ($\sim2.5\,\rm AU$). For the ray-tracing, the disks are assumed to be at a distance of 150 pc.}
\label{fig:rings}
\centering
\begin{subfigure}{0.49\textwidth}
  \centering
  \includegraphics[width=\linewidth]{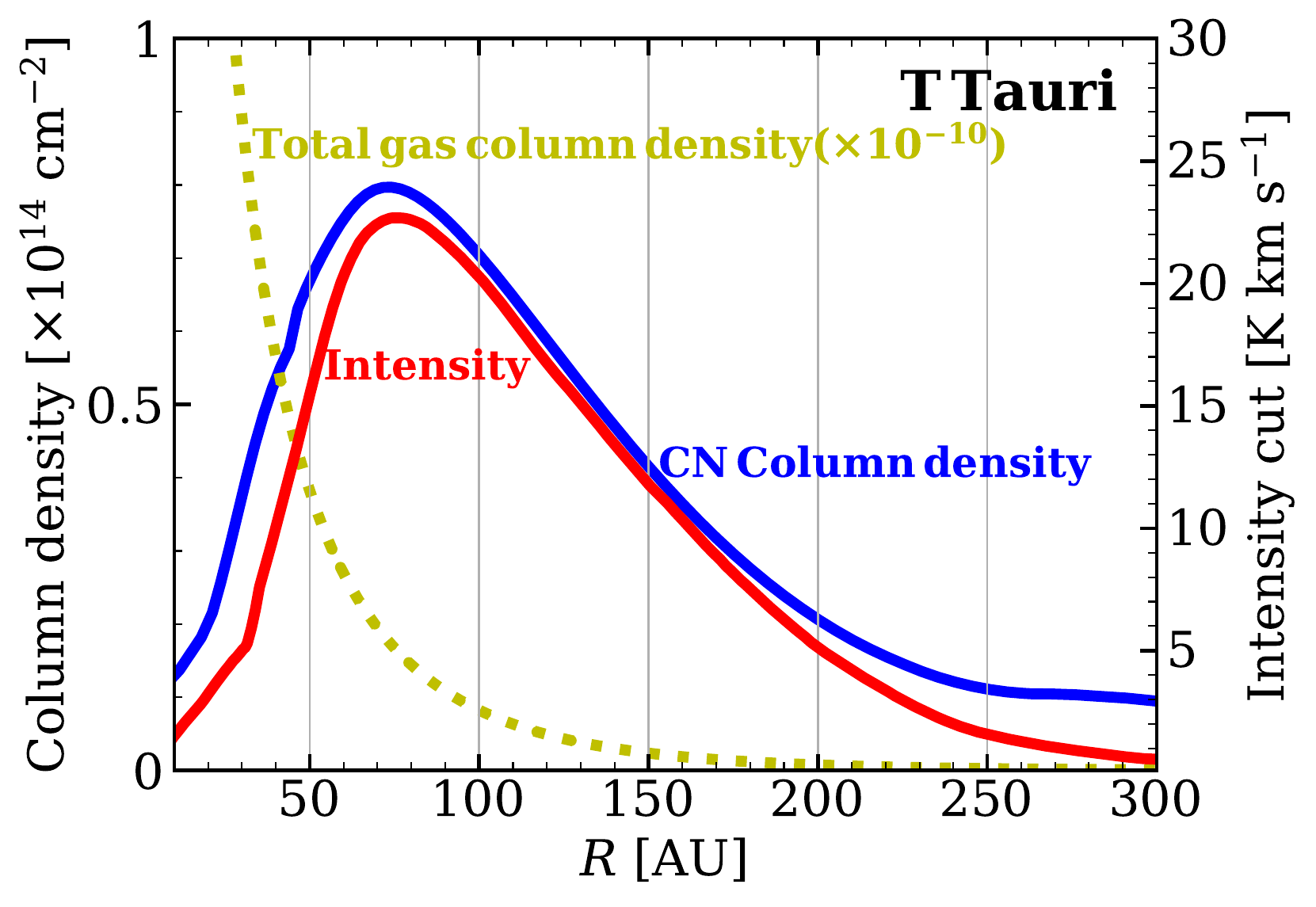}
  \phantomcaption
\end{subfigure}
\begin{subfigure}{.49\textwidth}
  \centering
  \includegraphics[width=\linewidth]{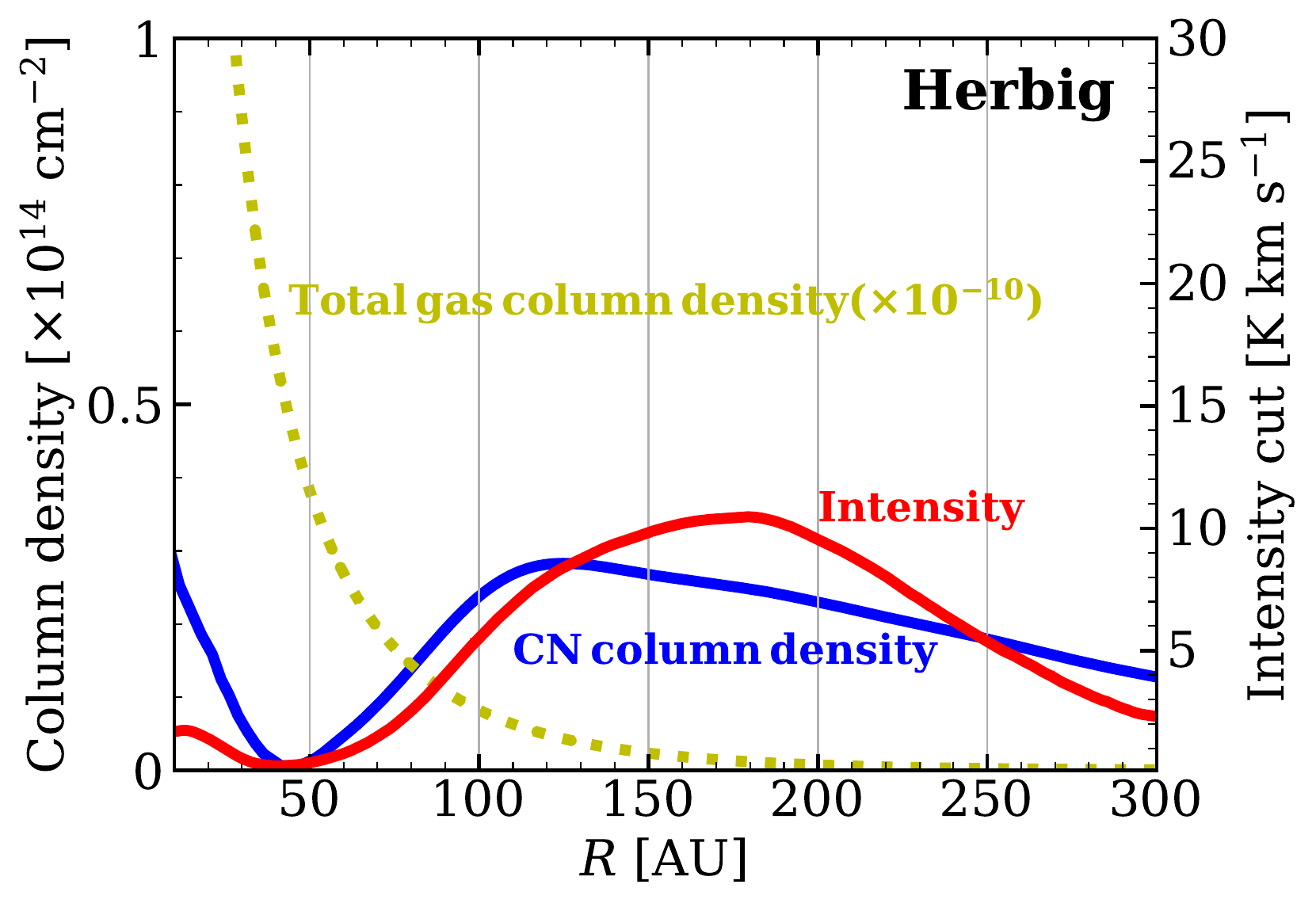}
  \phantomcaption
  \end{subfigure}
\caption{Comparison between the radial CN column density profile (blue) and intensity profile (red) for the two disks shown in Fig. \ref{fig:rings}. The total gas  column density multiplied times $10^{-10}$} is also plotted in yellow as a reference.
\label{fig:columndens}

\centering
  \includegraphics[width=\linewidth]{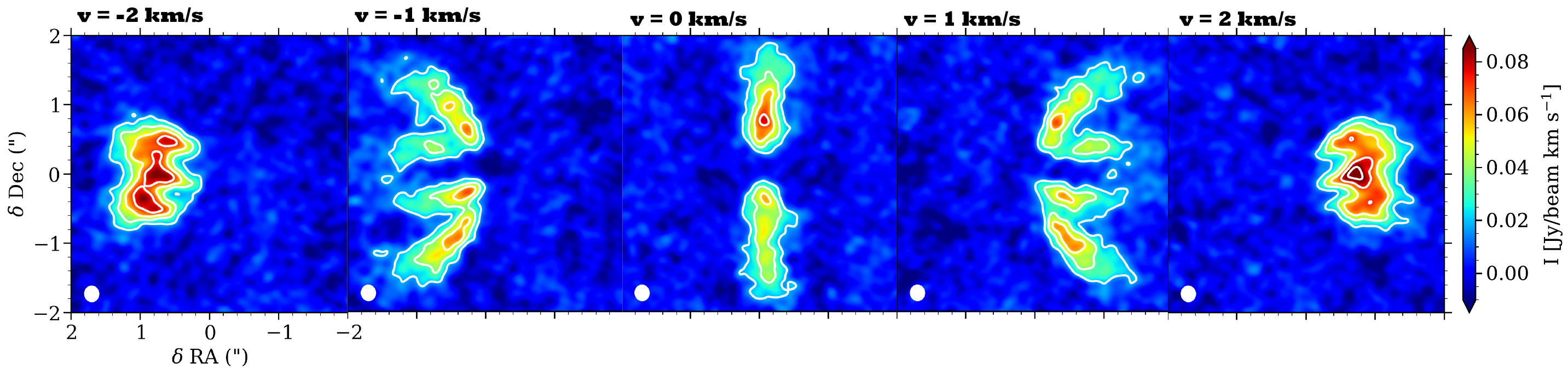}
\caption{Simulated observations of a $M_{\rm disk}=10^{-2}\, M_{\odot}$, $h_{\rm c}=0.1$, $\psi=0.2$ disk around a Herbig star, inclined of $60^\circ$. The image was simulated using the tasks \texttt{simobserve} and \texttt{simanalyze} in the CASA software.}\label{fig:ch_map}
\end{figure*}
\clearpage
}

\section{Results}\label{sec:results}

\subsection{CN abundance}\label{sec:cnabu}

CN is often referred to as a PDR tracer, and many studies have shown
that in protoplanetary disks this molecule forms and survives mostly
in the surface layers exposed to intense UV radiation. The same feature is found in our models, with the
highest CN abundance region close to the surface and outer edge
of the disks.

Fig.~\ref{fig:cnabu} shows the CN abundance (with respect to the total gas density) for a typical T Tauri disk. Highest CN abundances are
located high up in the disk between $10$ and $100\,\rm AU$ and extend
down to the midplane at the disk edge at large radii where the external UV also plays a role.  Such an abundance distribution in turn translates into a
ring of high column densities (Fig. \ref{fig:cncolumndens}), which is consistent
with a ring-shaped emission intensity profile (see
Sec. \ref{sec:emission}). In contrast, the HCN abundance and column
density peak at small radii  (see Fig.~\ref{fig:cncolumndens}).

Fig. \ref{fig:Vertical_cuts}. shows a vertical cut of the abundances of the main species at
the radial location of the CN abundance maximum (green dot in Fig. \ref{fig:cnabu}). Again, the CN molecule is located near the surface
of the disk and between the hydrogen atomic and molecular hydrogen layer, with HCN located deeper into the disk.

It is worth investigating whether the increase in CN column density
towards the outer disk is due to decreasing UV, decreasing
temperature, or some combination of parameters.  The strongest
dependence on temperature is found in the back-reaction
\begin{equation}
\rm CN+H_2\rightarrow HCN+H,
\end{equation}
which has an activation temperature of $900\,\rm K$
\citep{baulch2005}. However, while this reaction is important in
increasing HCN in the inner disk, it does not play a role in
setting the CN abundance since the newly formed HCN is readily
photodissociated back to CN. 

\begin{figure}[]
\centering
\begin{subfigure}{0.49\textwidth}
  \centering
  \includegraphics[width=\linewidth]{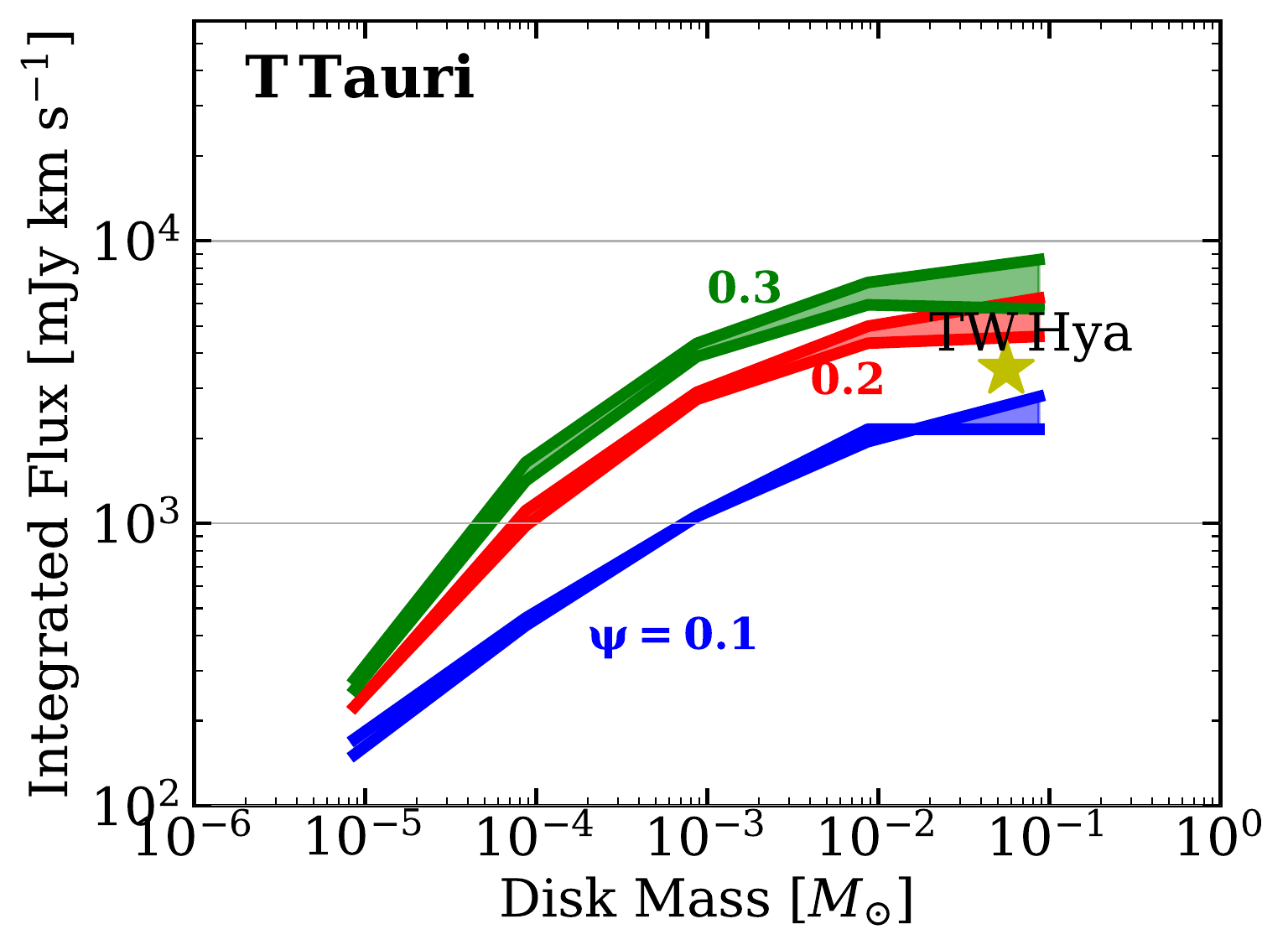}
  \phantomcaption\label{fig:ttau_flux_mass}
\end{subfigure}
\begin{subfigure}{.49\textwidth}
  \centering
  \includegraphics[width=\linewidth]{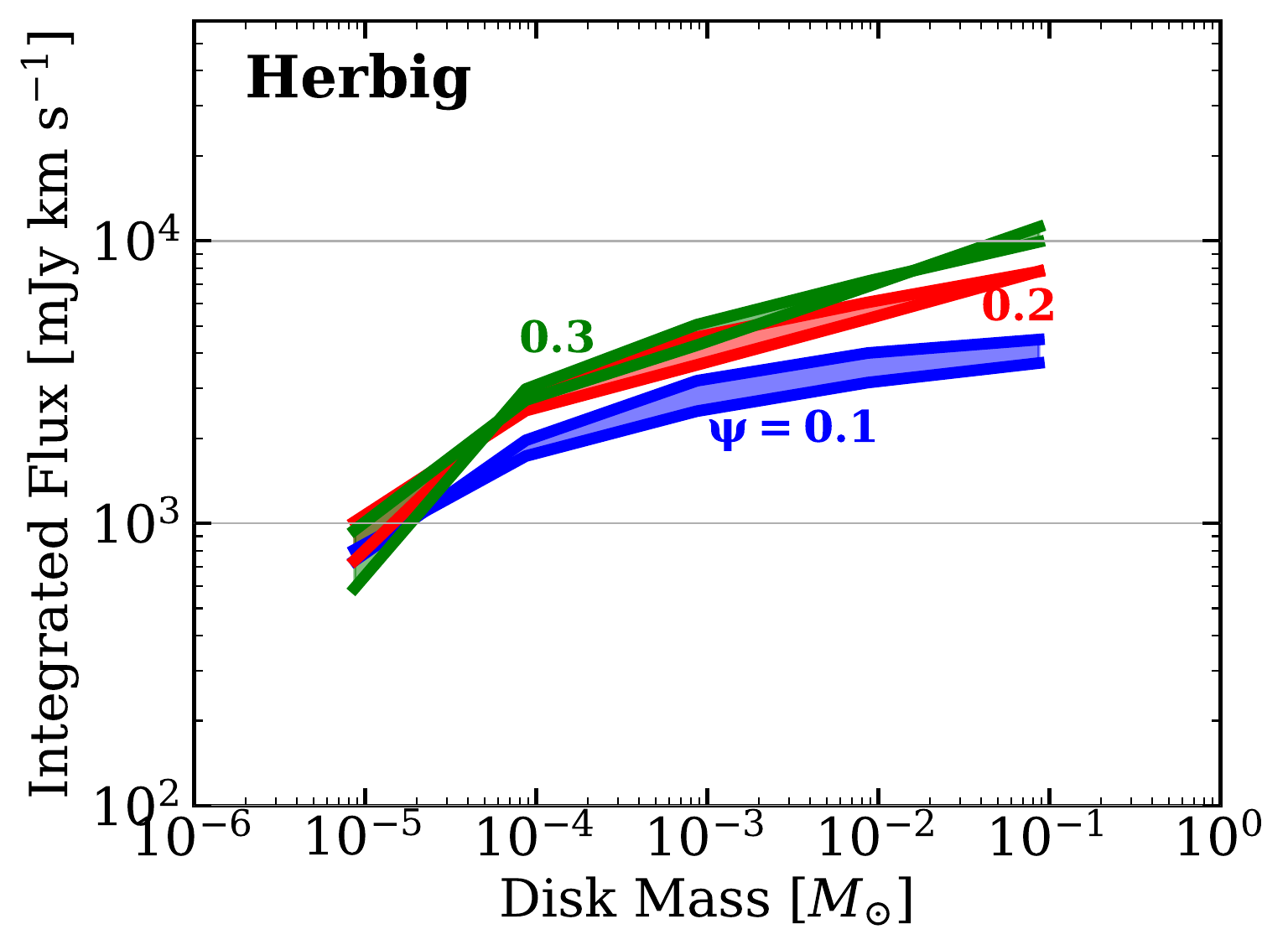}
  \phantomcaption\label{fig:herbig_flux_mass}
\end{subfigure}
\caption{Dependence of the CN integrated flux on the total disk mass for disks around T Tauri (upper panel) and Herbig (lower panel) stars. For each value of $\psi$, the range highlighted by the shaded area is due to the variation of $h_{\rm c}$. The fluxes are measured by assuming the disks are at a distance of 150 pc. The value measured for TW Hya and rescaled for the distance of the models is also shown.}
\label{fig:flux_mass}
\end{figure}

The main reactions that dominate the formation of CN are instead
\begin{eqnarray}
\rm N+H_2^* \rightarrow &\rm NH+H\label{eq:nhform} \\
\rm C^++NH \rightarrow &\rm CN^++H. \label{eq:cnplusform}
\end{eqnarray}
CN$^+$ proceeds then to CN through a charge transfer reaction with H
or through the intermediates HCN$^+$ and HCNH$^+$ followed by
dissociative recombination.  Alternatively, NH can react with neutral
C directly to form CN.  Another less important route to cyanides
starts from reactions of N with CH$_2$ and C$_2$H. Alternatively,
reactions of N$^+$ with H$_2$ lead to NH$_4^+$, through a series of ion-molecule reactions, which then reacts
with C to form HCNH$^+$.

CN is primarily destroyed by photodissociation and by reactions with
atomic O. The rate coefficient of the CN + O reaction is taken to have
no temperature dependence, as discussed in \citet{visser2016} and
consistent with the KIDA recommendations.

\begin{figure}
\begin{subfigure}{0.49\textwidth}
  \centering
  \includegraphics[width=\linewidth]{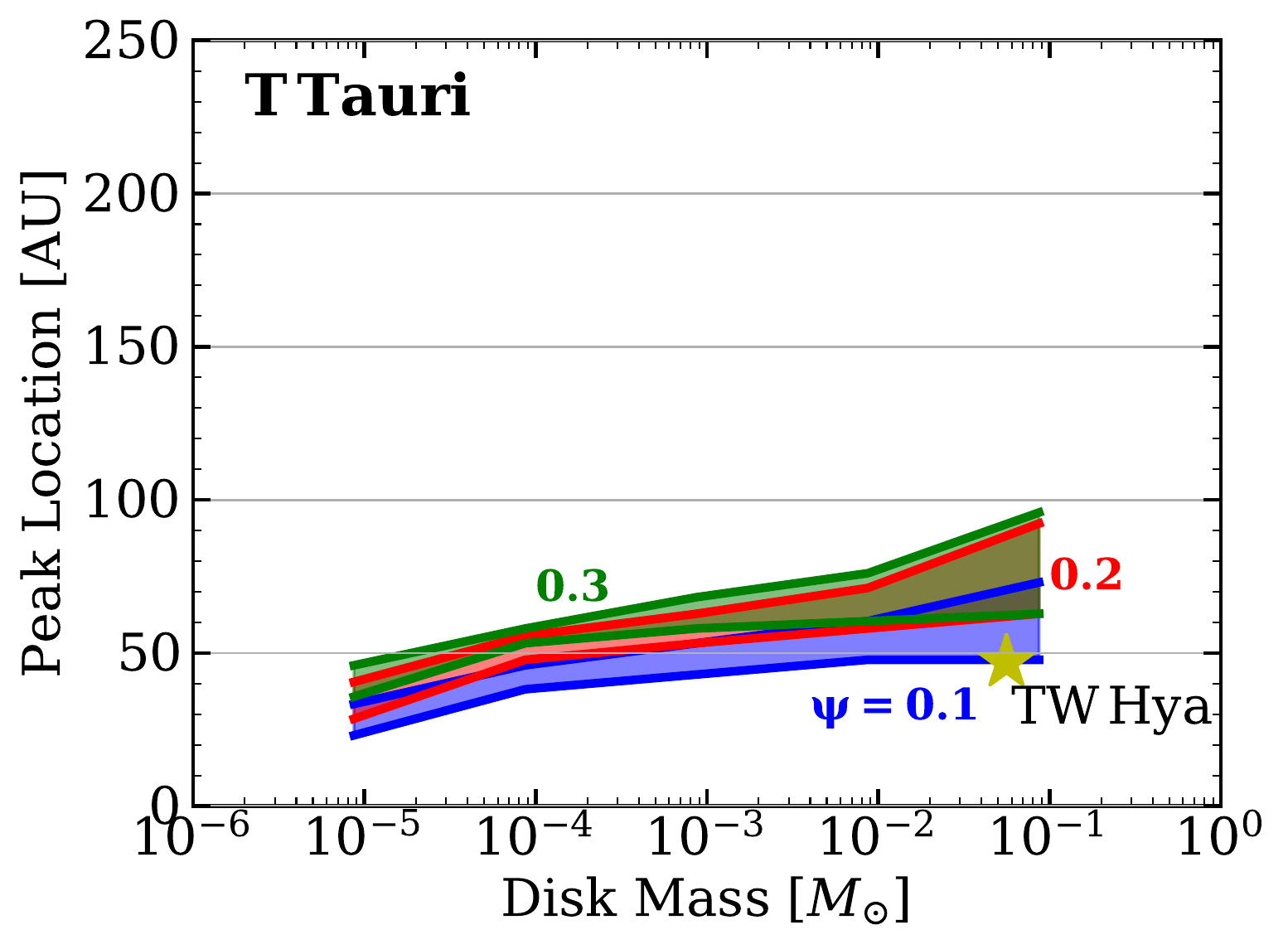}
  \phantomcaption\label{fig:ttau_ring_mass}
\end{subfigure}
\begin{subfigure}{.49\textwidth}
  \centering
  \includegraphics[width=\linewidth]{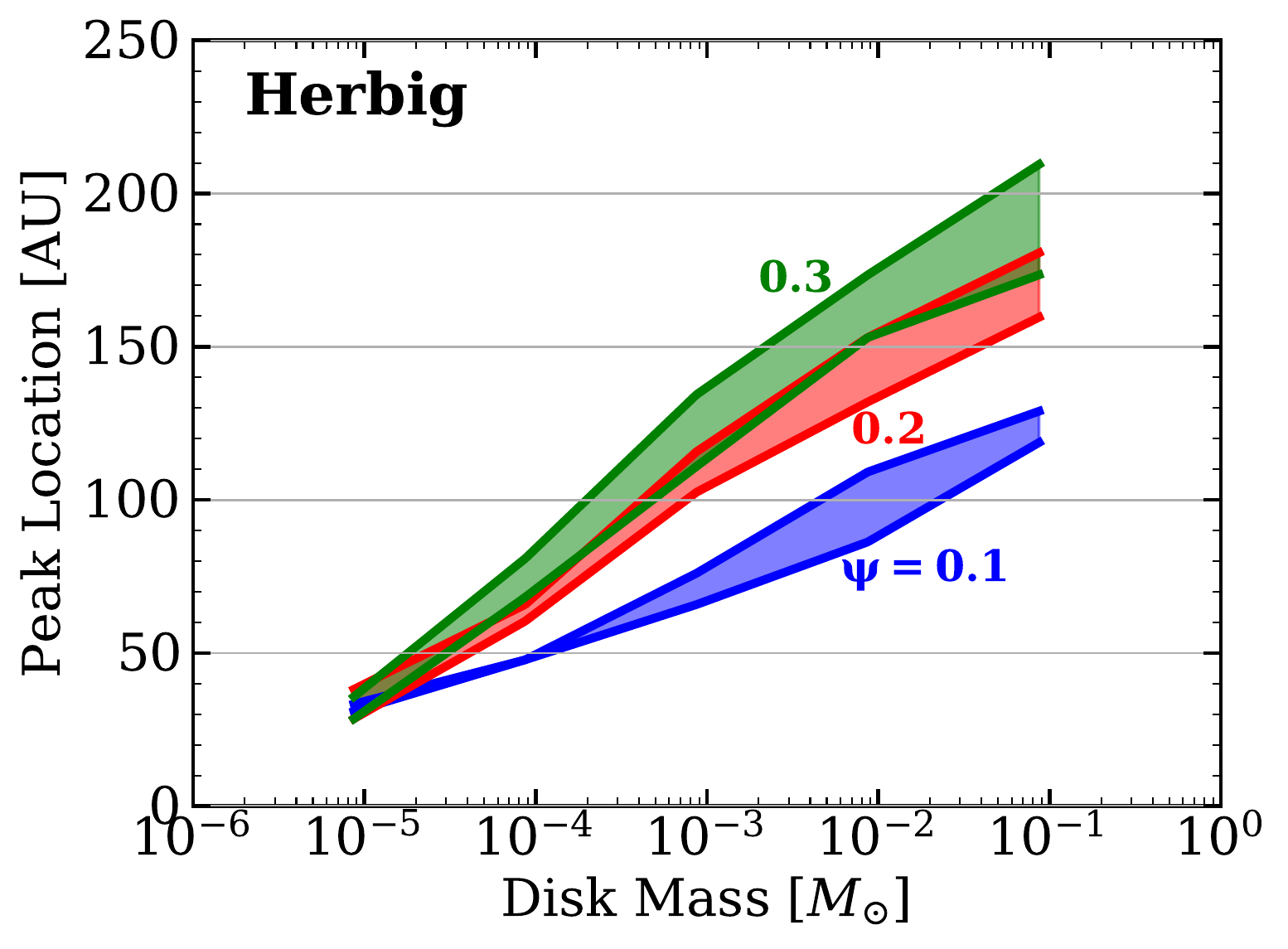}
  \phantomcaption\label{fig:herbig_ring_mass}
\end{subfigure}
\caption{Dependence of the CN ring location on the total disk mass in the models.}
\label{fig:ring_mass}
\end{figure}

The normalized abundances of the most relevant species are shown in
Fig. \ref{fig:radial_abus} along the dotted green line shown in Fig.~\ref{fig:cnabu}. The peak of the CN abundance
corresponds to that of NH. In turn, H$_2^*$ is required in
order for NH to exist. NH also shows a second inner peak at 20 AU. This, however, does not translate into a second CN peak because of the lack of C$^+$ in this region, which is required for reaction (\ref{eq:cnplusform}) to happen. The net
result is that outside 30 AU the normalized abundance profiles of H$_2^*$ , NH, and CN are
qualitatively similar.

The H$_2^*$ abundance is controlled by the molecular hydrogen
excitation and de-excitation rates shown in Fig. \ref{fig:h2_cd_exc} . Moving inward from the CN abundance peak, the
de-excitation of H$_2^* $ increases faster than the excitation due to
FUV pumping; this leads to a decrease in $x$(H$_2^*$), and consequently also in $x$(NH)
and $x$(CN). At larger radii the FUV radiation field is
attenuated by the dust, de-excitation turns from collisional to
spontaneous emission and starts to dominate again, leading to a decrease in
$x$(H$_2^*$) and x(CN).  Their ratio (solid red line in Fig. \ref{fig:h2_cd_exc}) thus has its maximum right at the location of the CN abundance maximum.

In general, the dependence of CN abundance on the balance between H$_2$ excitation (via FUV pumping) and de-excitation (via collisions) can be summarized with the local ratio between FUV flux $G_0$ and gas density $n_{\rm gas}$.  In fact, FUV pumping is proportional to $G_0\times n_{\rm gas}$, while collisional de-excitation scales as $n_{\rm gas}^2$.  We observe that the value of $G_0$/$n_{\rm gas}$  at the CN abundance peak is constant for all of our models; this is consistent with the UV controlling the abundance of CN. When $G_0$/$n_{\rm gas}$ is too low, collisional de-excitation dominates over FUV pumping, thus hindering the formation of H$_2^*$. On the other hand, when $G_0$ is high enough, H$_2$ gets photodissociated. Photodissociation scales with $G_0\times n_{\rm gas}$ and formation with $n_{\rm gas}^2$. If $G_0/n_{\rm gas}$ is too high, photodissociation dominates over formation, thus removing molecular hydrogen and preventing the formation of H$_2^*$. The H$_2^*$ layer is located between the molecular and atomic hydrogen layers, and therefore the maximum of H$_2^*$ and consequently of CN is located where the correct ratio between $G_0$ and $n_{\rm gas}$ is met (see Fig. \ref{fig:g0ngas} in the Appendix).

Finally, previous astrochemical models have shown that X-rays can play an important role in the chemistry of CN \citep{1996A&A...306L..21L}. However, \citet{2005A&A...440..949S,2007A&A...466..977S} find that the effect of X-rays in negligible in environments where a high FUV flux is present. Our results indeed do not show significant differences when the $L_{\rm X}$ parameter is varied between $10^{27}$ and $10^{31}\,\rm erg\,s^{-1}$.

\subsection{Emission}\label{sec:emission}

\begin{figure}

  \includegraphics[width=\linewidth]{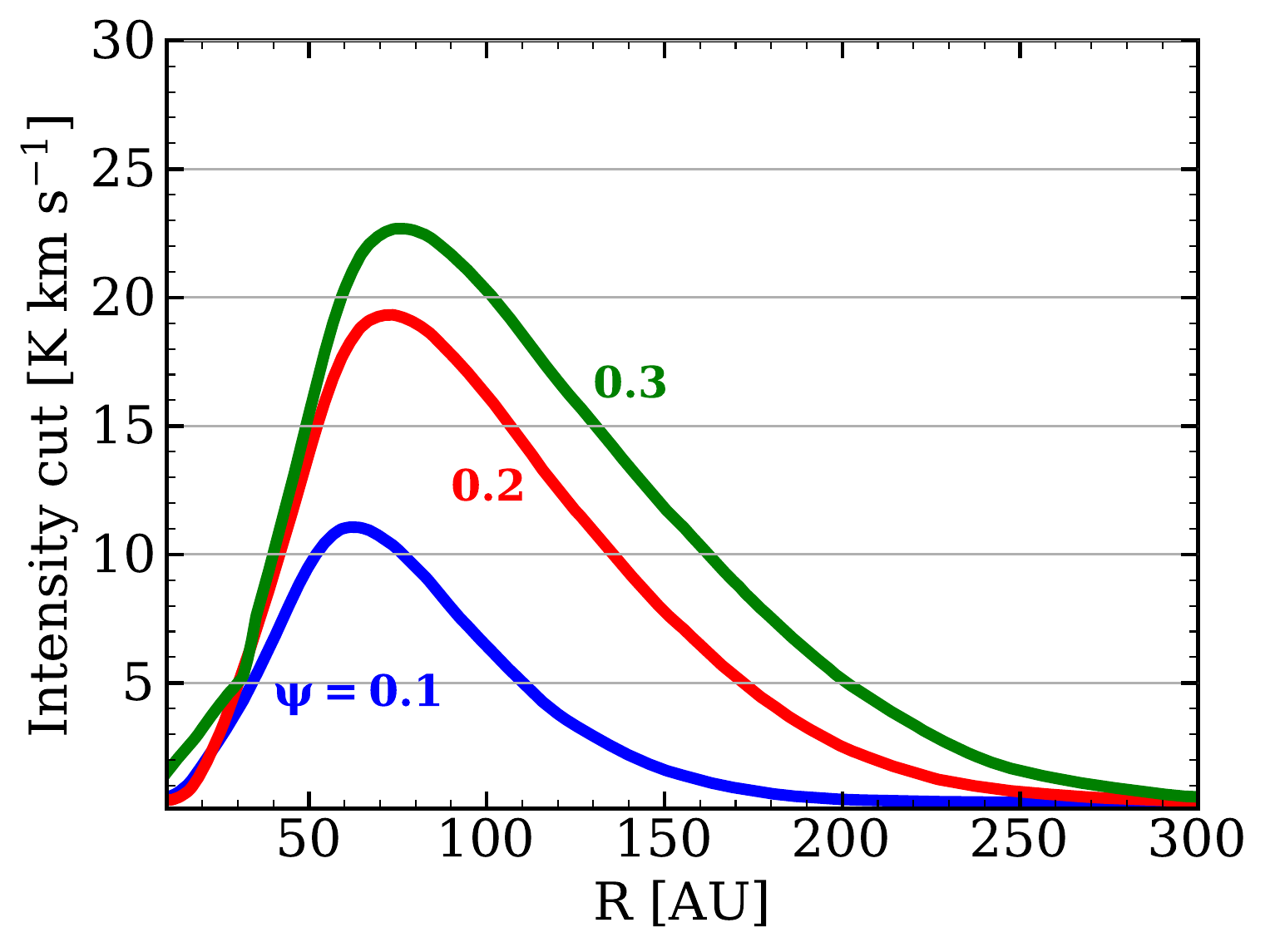}

\caption{Different intensity radial profiles for $\psi=0.1,0.2$ and $0.3$ in a disk with $M_{\rm disk}=10^{-2}\, M_{\odot}$ and $h_{\rm c}=0.1$ rotating around a T Tauri star. The intensity of the CN  emission increases in more flared disks and moves to slightly larger radii.
}\label{fig:psicomp_int}
\end{figure}

\begin{figure}
\begin{subfigure}{0.49\textwidth}
\centering
  \includegraphics[width=\linewidth]{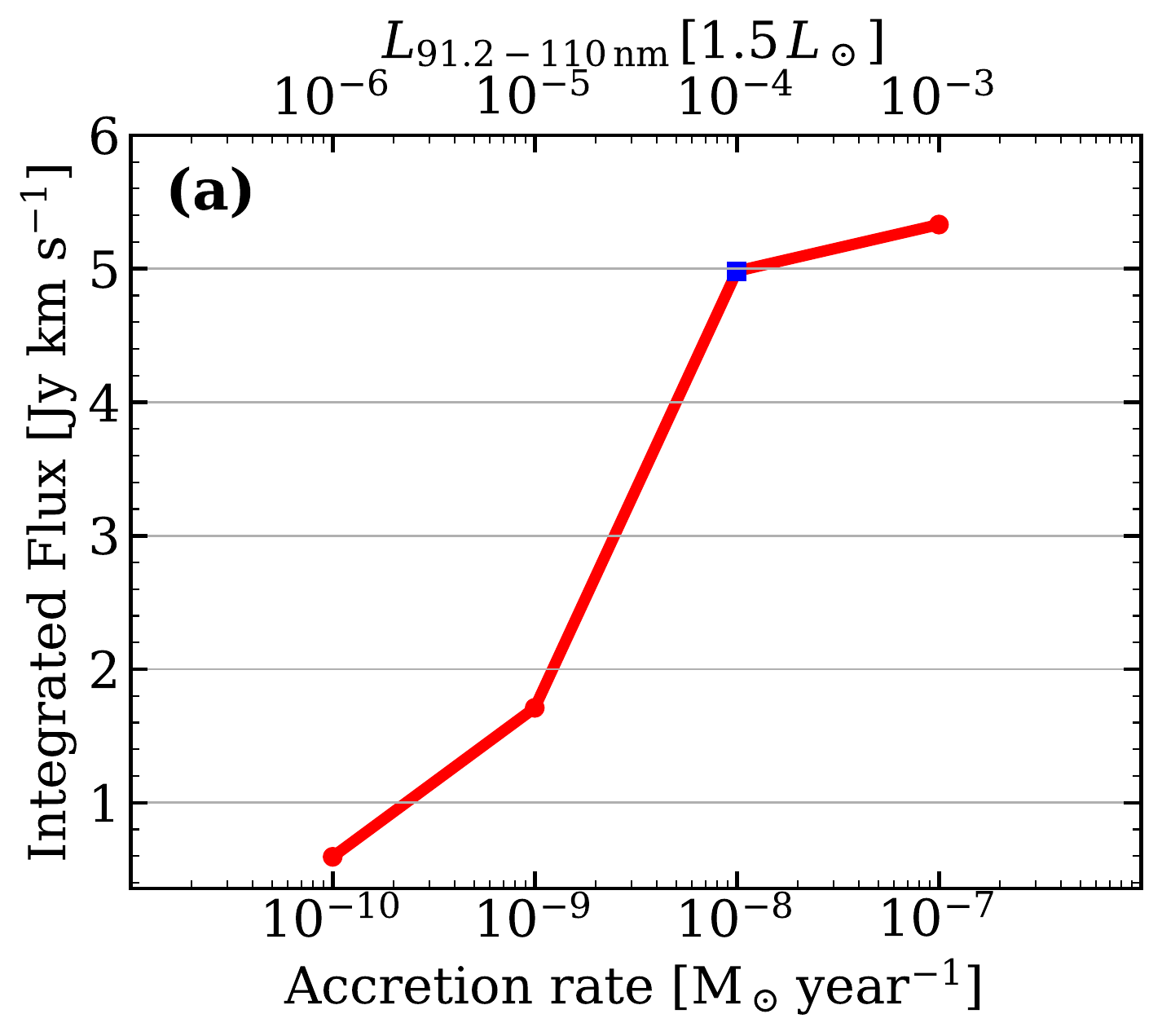}
\phantomcaption\label{fig:fluxes_accretions}
\end{subfigure}
\begin{subfigure}{0.49\textwidth}
\centering
  \includegraphics[width=\linewidth]{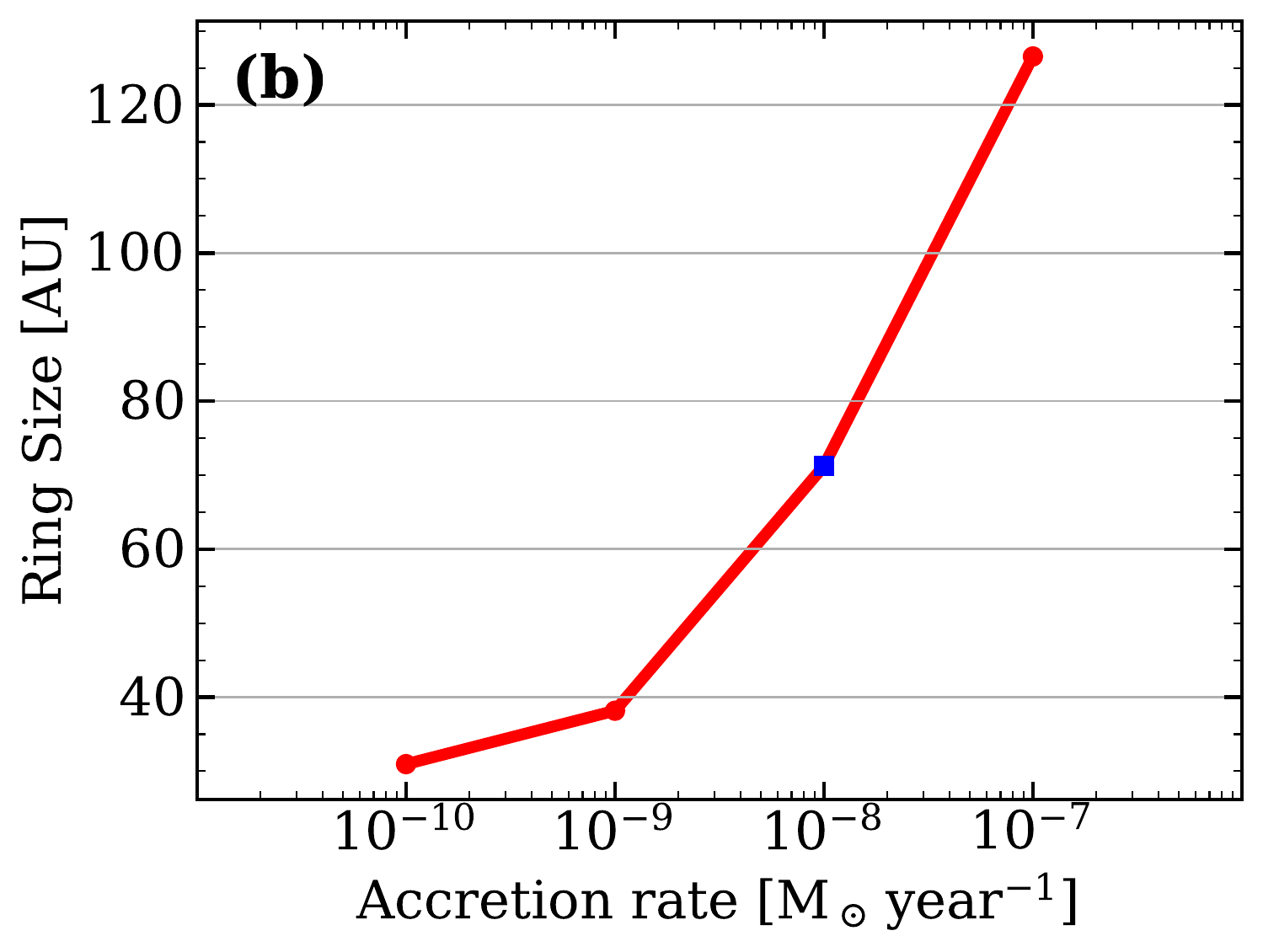}
  \phantomcaption\label{fig:rings_accretions}
  \end{subfigure}
\caption{(a) Integrated flux of the CN ($N,J$)=(3-2,7/2-5/2) emission in a T Tauri disk  with $M_{\rm gas}=10^{-2}\,M_\odot$, $\psi=0.2$ and $h_{\rm c}=0.1$ as a function of the accretion rate on the central star. For each accretion rate, the UV luminosity between 91.2 and 110 nm is also shown at the top of the figure. The standard model (\mbox{$\dot M=10^{-8}$ M$_\odot$ year$^{-1}$}) is marked with a blue square. (b) Same as (a), for the CN ring radius.}\label{fig:acc_trends}
\end{figure}

From the temperature structure and chemical abundances obtained for each model it is possible to ray-trace CN emission. In the CN molecule, each rotational energy level (labelled with the quantum number $N$) is split into a doublet by the spin-rotation interaction (quantum number $J$), thus leading to three fine structure transitions. Each of the two sub-levels is in turn split into a triplet by nuclear magnetic interactions (quantum number $F$), giving rise to hyperfine transitions. In this work we ignored the latter transitions, and discuss only the fine structure transitions. In particular, if not otherwise specified, on the brightest of the three, namely the CN ($N,J$)=(3-2,7/2-5/2) transition at 340.25 GHz. Fig. \ref{fig:rings} shows the ray-traced images for two disks from our grid that have the same physical structure but rotate around a T Tauri and a Herbig star, respectively. In the ray tracing, the disks are set to be face-on to allow the CN emission to be clearly distinguishable. The distance of the disks is assumed to be 150 pc. 

\begin{figure}
\begin{subfigure}{0.49\textwidth}
\centering
  \includegraphics[width=\linewidth]{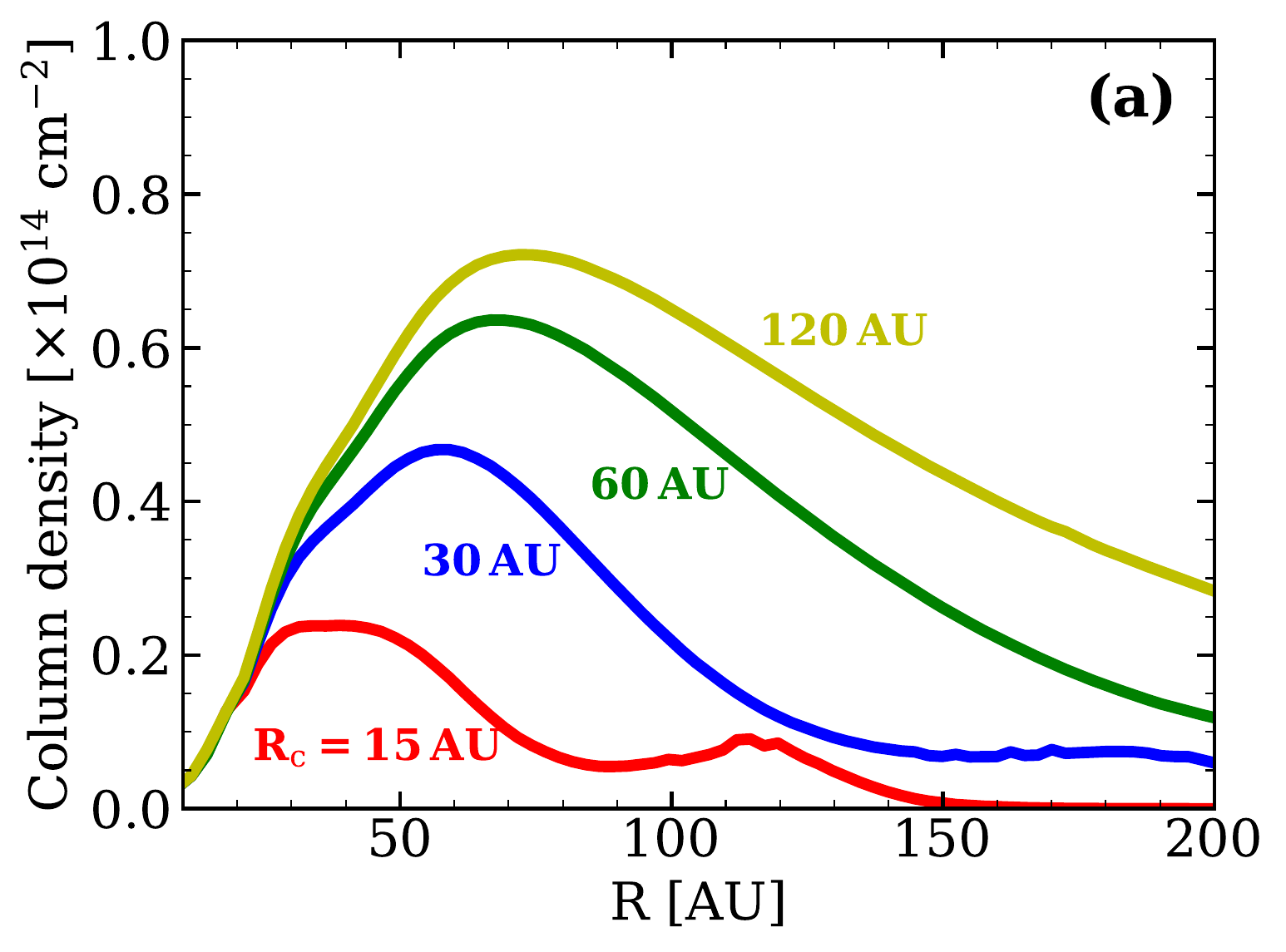}
\phantomcaption\label{fig:cd_rcs}
\end{subfigure}
\begin{subfigure}{0.49\textwidth}
\centering
  \includegraphics[width=\linewidth]{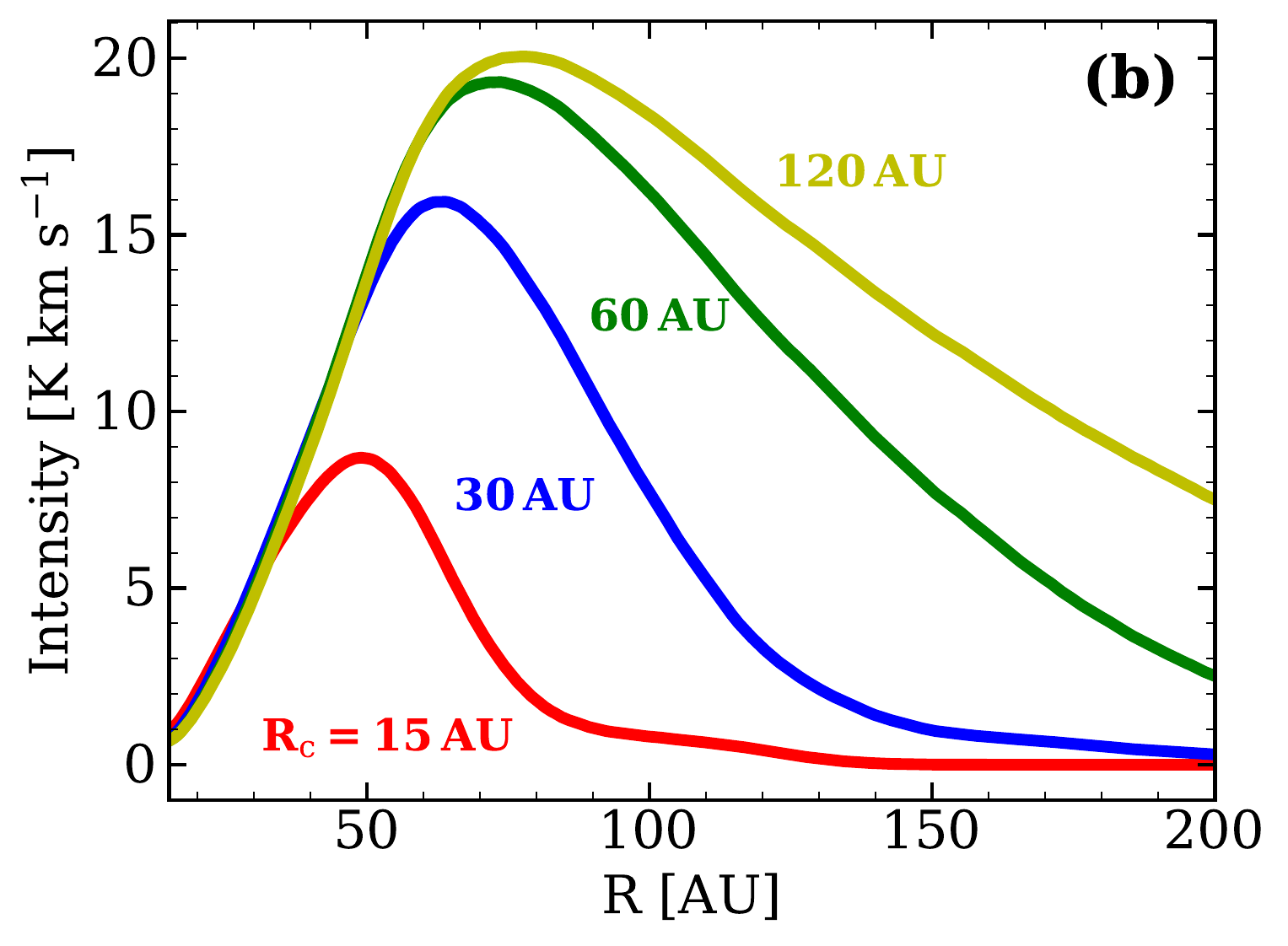}
  \phantomcaption\label{fig:int_rcs}
  \end{subfigure}
\caption{Comparison between (a) column density radial profiles and (b) radial intensity cuts  in models with different $R_{\rm c}$ values.}
\end{figure}

\begin{figure}
\begin{subfigure}{0.49\textwidth}
\centering
  \includegraphics[width=\linewidth]{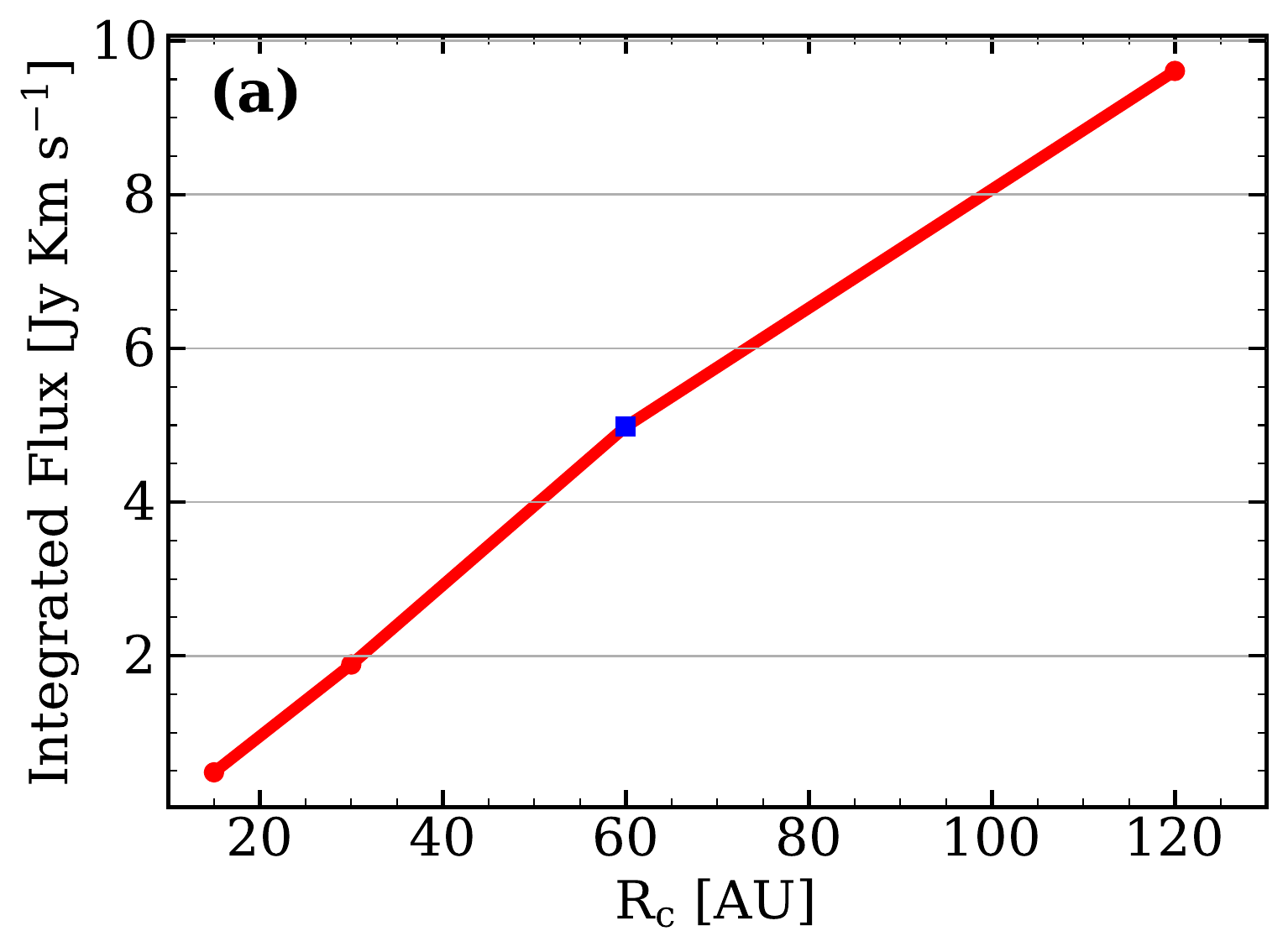}
\phantomcaption\label{fig:fluxes_rcs}
\end{subfigure}
\begin{subfigure}{0.49\textwidth}
\centering
  \includegraphics[width=\linewidth]{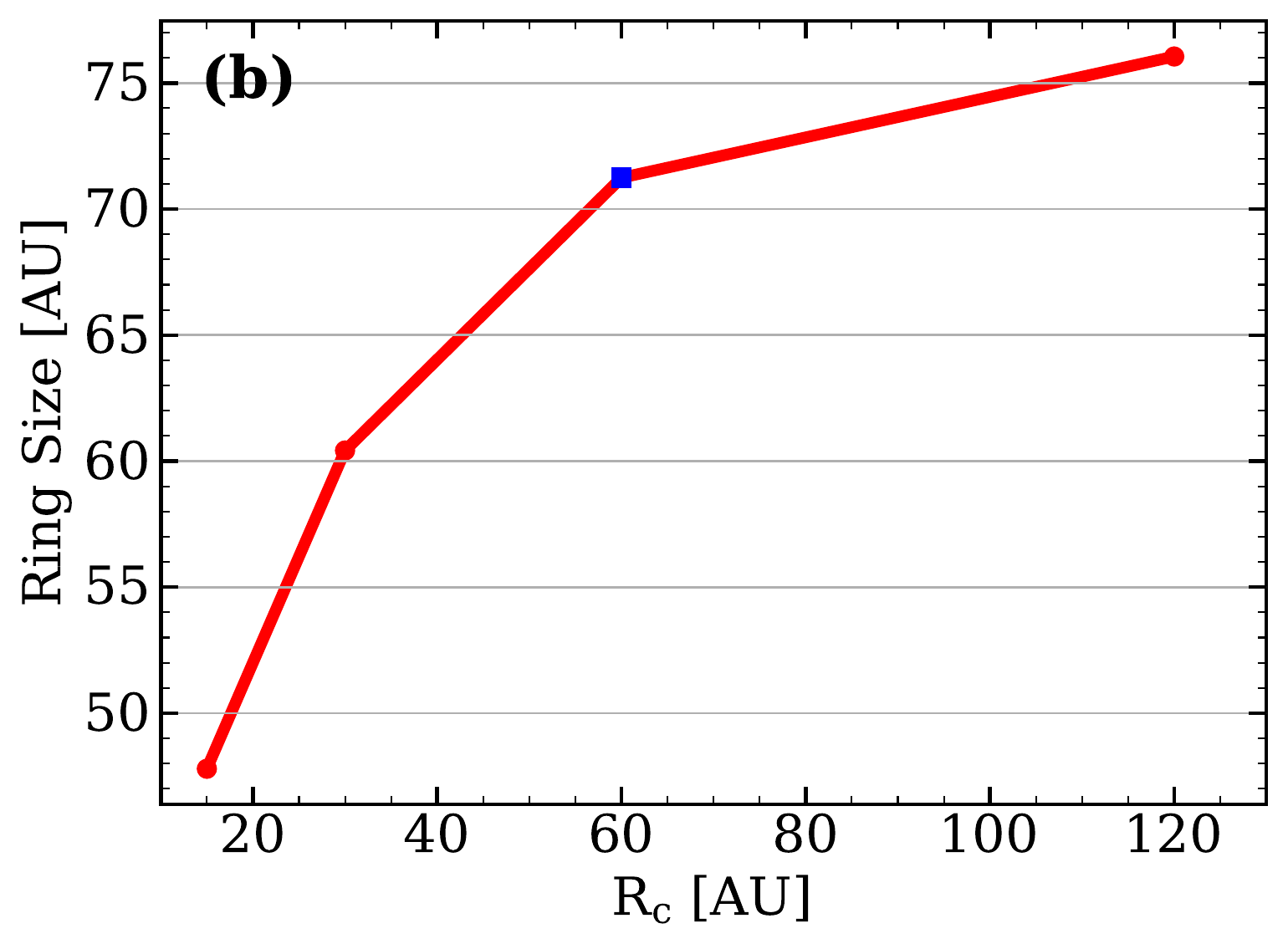}
  \phantomcaption\label{fig:rings_rcs}
  \end{subfigure}
\caption{Dependence of (a) integrated fluxes and (b) ring size  in models with different $R_{\rm c}$ values. The standard model (\mbox{$R_{\rm c}=60 \,\rm AU$}) is marked with a blue square.}
\end{figure}

As shown in Fig. \ref{fig:cnabu}, CN is optically thin. Accordingly, for T Tauri stars the radial intensity profile is consistent with the radial column density profiles, and the peak of the column density is located at the same radius as that of the emission peak (Fig. \ref{fig:columndens}). In the disks around Herbig stars, CN forms at larger radii than in T Tauri stars. In addition, in some Herbig models, such as that shown in Figs. \ref{fig:rings} and \ref{fig:columndens}, emission can peak at even larger radii than the column density peak. This is because the peak in overall CN column density does not correspond to the peak in column density of the CN molecules in the upper state of the transition we are studying (N=3, J=7/2), which is located further out. A detailed explanation of this excitation mechanism can be found in Appendix \ref{app:levelpop}.

All other models in our study show similar ring-like structures, thus confirming that ring-like emission is an intrinsic feature of CN emission in full disks. In every case, such a structure is due to the higher abundance of CN at larger radii and is not caused by ring-shaped features in the global disk morphology such as those characteristic in transitional disks.

We also concluded that it is not due to sub-thermal excitation effects, which in previous studies have been shown to potentially play an important role for high dipole moment molecules such as CN  \citep[e.g., ][]{vz2003}, confirming the conclusions of \citet{2017A&A...603L...6H} for the TW Hya disk. Nevertheless we find that the high critical density of CN does play a role in enhancing the size of CN rings, especially in Herbig stars.

Finally, given the fact that CN in our models is mostly abundant in the uppermost layers of the disk, CN can be used as a probe for the vertical structure of disks. Fig. \ref{fig:ch_map} shows a simulated ALMA observation of CN in an inclined disk around a Herbig star,  for five representative channels 0.25 km/s wide: even with a modest resolution of 0.22" and with 30 minutes of on-source integration, the two layers are clearly distingushable.

\subsection{Dependence on mass and flaring}\label{sec:parameters}
We investigated how CN emission is affected by some of the key parameters of our models. First, we studied the dependence of the integrated flux and the location of the ring on the total gas mass of the disks.

Fig. \ref{fig:flux_mass} shows that both for Herbig and T Tauri stars the CN integrated flux increases as the total disk mass increases.  The measured fluxes range from $200\,\rm mJy\, km\,s^{-1}$ and $9\,\rm Jy\, km\,s^{-1}$, which is consistent with observed values \citep[e.g. ][]{guilloteau2013}. The dependence of the flux on the mass is steeper at low masses for T Tauri stars and flattens for $M_{\rm disk}>10^{-3}\,\rm M_{\odot}$. For the Herbig stars, the flux generally increases less steeply with mass. For both stellar spectra, more flared disks always show higher intensities.

Fig. \ref{fig:ring_mass}, on the other hand, shows the size of the ring as a function of $M_{\rm disk}$. More massive disks have larger rings. The difference between T Tauri and Herbig stars is much more evident here than for the flux, hinting at a strong dependence of the CN abundance on the FUV spectrum.  In particular, disks surrounding T Tauri stars show rings with sizes ranging between $30\,\rm AU$ and $100\,\rm AU$ while Herbig disks show rings extending to more than $200\,\rm AU$. The disk scale height, parametrized through $h_{\rm c}$, does not play a strong role, as the scatter due to the variation of $h_{\rm c}$ is small compared to the ranges of the fluxes and of the ring sizes, which from now on we define as the radial location of the emission peak.

\begin{figure}
\centering
  \includegraphics[width=\linewidth]{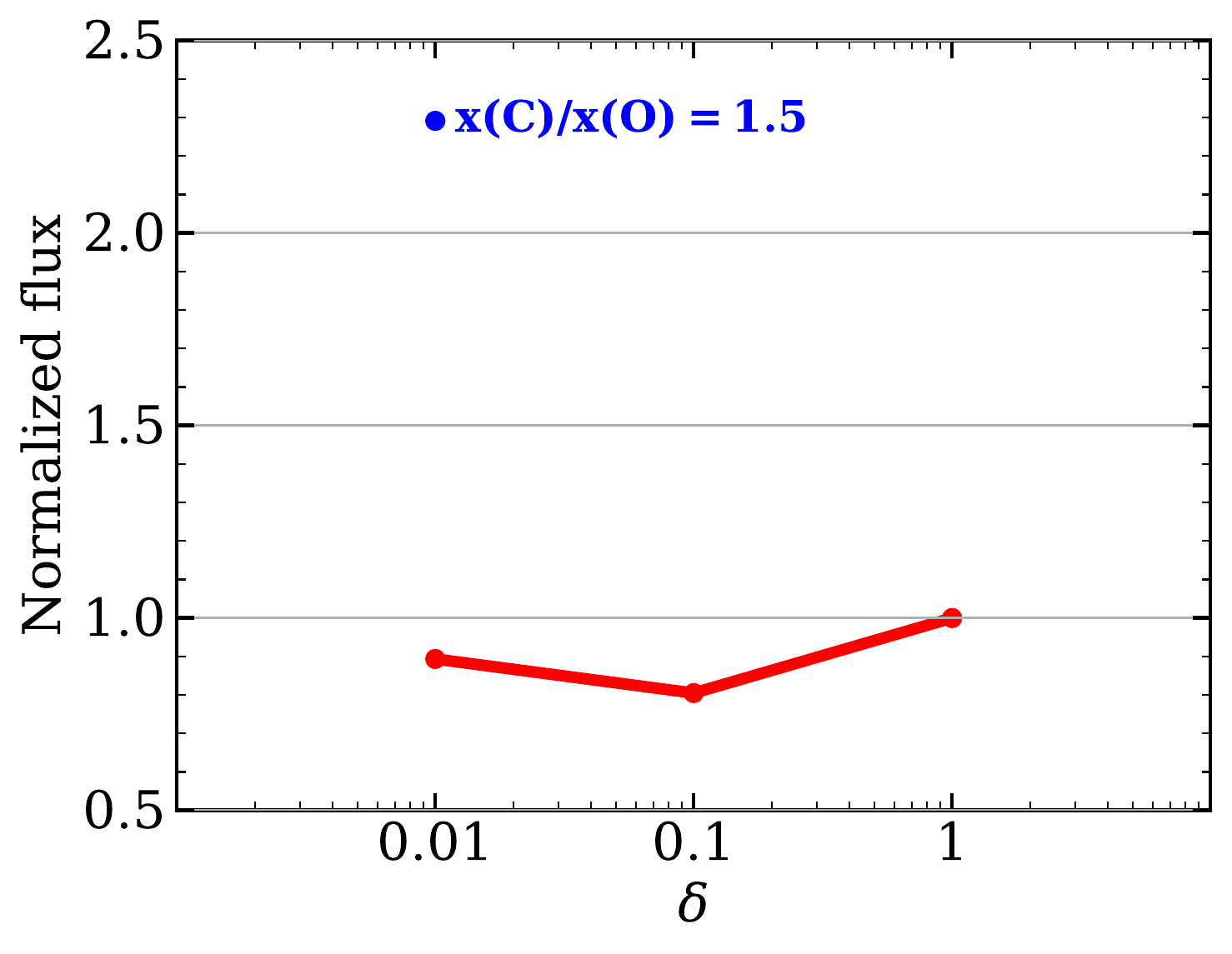}

\caption{Integrated flux of the CN ($N,J$)=(3-2,7/2-5/2) emission in a disk  with $M_{\rm gas}=10^{-2}\,M_\odot$, $\psi=0.2$ and $h_{\rm c}=0.1$ as a function of the C and O depletion. The blue dot refers to the model where a higher gas-phase carbon abundance than oxygen is assumed. All the fluxes are normalized to the non-carbon-depleted $\delta=1$ case. The fluxes vary by less than a factor of 2.}\label{fig:fluxes_depletions}
\end{figure}

Finally, disks with a larger flaring angle, exposed to stronger FUV fluxes, show systematically larger rings for both Herbig and T Tauri stars. Changing the flaring angle of the disks in our models has an impact on the CN abundance and consequently on its emission intensity. Fig. \ref{fig:psicomp_int} shows a model with  $M_{\rm disk}=10^{-2}\,\rm M_{\odot}$ and $h_{\rm c}=0.1$ rotating around a T Tauri star as an example. The radial intensity cuts are shown for the three different values of $\psi$ used in our grid. As $\psi$ increases, the emission increases in the ring, which moves outward. This confirms our earlier conclusion that UV flux plays an important role in regulating the CN abundance. 
For each model, the vertical location of the CN peak is determined by the UV field relative to the gas density (Fig. \ref{fig:g0ngas}):  when the flaring angle is larger the optimal ratio between UV flux and gas density is found at lower altitudes where the gas density is higher.

The same qualitative trends are observed in our models for the \mbox{$(N,J)=(3-2,5/2-5/2)$} and \mbox{$(N,J)=(3-2,5/2-3/2)$} fine transitions and for the brightest of the $N=2-1$ transitions, namely the $J=5/3-3/2$ one at 226.88 GHz. Finally, the ratio between the integrated flux of the $(N,J)=(3-2,7/2-5/2)$ transition and that of the $(N,J)=(2-1,5/2-3/2)$ one lies between 0.5 and 3 in all our models.

Finally the dependence of the CN emission pattern on the settling and degree of growth of the dust grains was studied. For this purpose, two additional models were run of a $10^{-2}\,\rm M_{\odot}$, $\psi=0.2$, $h_{\rm c}=0.1$ disk surrounding a T Tauri star  assuming $f_{\rm large}=0.1$ and 0.8. In these models, the higher amount of small grains in the disk atmosphere lowers the intensity of the FUV field. Accordingly, the high CN-abundance region is moved to even higher altitudes in the disks in which $G_0/n_{\rm gas}$ is similar. As a consequence, the CN column density and integrated flux decreases by a factor of $\sim 3$ for the $f_{\rm large}=0.8$ model, and the rings are only a few AUs smaller than in the $f_{\rm large}=0.99$ case. In the extreme $f_{\rm large}=0.1$ model, however, the column density and intensity decrease by almost an order of magnitude.

\subsection{Excess UV from accretion}
In order to further test the impact of the UV field on the CN flux and distribution, we ran an additional set of models by varying the excess UV due to accretion onto a T Tauri star. So far, the T Tauri stellar spectrum was modelled as a 4000 K black-body spectrum with a UV excess for an accretion rate of $10^{-8}\, M_\odot\rm year^{-1}$. The excess accretion UV was modelled as a 10000 K black body on a $1.65\, R_\odot$ radius star  \citep{kama2016}.   The additional accretion rates considered are $10^{-7}$, $10^{-9}$, $10^{-10}\, M_\odot\rm year^{-1}$ and a non-accreting black-body spectrum, and we investigated the effects of the different UV excess on a disk with $M_{\rm gas}=10^{-2}\,M_\odot$, $\psi=0.2$ and $h_{\rm c}=0.1$.

The integrated flux and ring size as functions of accretion rate are plotted in Fig. \ref{fig:acc_trends}. The observed trends presented in Fig. \ref{fig:acc_trends} confirm the main differences seen between T Tauri and Herbig stars. Higher accretion rates corresponding to higher UV fluxes show higher CN integrated emission. Moreover, the radius of the CN rings increases in size as the UV field becomes stronger. Of course, for a given accretion rate, a spread in the flux and ring size is expected due to the other parameters as explained above.

\subsection{Varying the disk size}
As a last test, we investigated the effect of different values of $R_{\rm c}$ on CN morphology, which ultimately reflects on the disk size. As a base model, we used the $M_{\rm gas}=10^{-2}\,M_\odot$, $\psi=0.2$, $h_{\rm c}=0.1$ T Tauri disk, and then varied $R_{\rm c}$ keeping the inner surface density profile equal in all the models (see Fig. \ref{fig:denscomp} in the appendix). In particular, we tested models with $R_{\rm c}=15,30,60$ and $120$ AU. Fig. \ref{fig:cd_rcs} shows that the inner part of the CN column density profiles remains constant in all models: this is consistent with the fact that CN abundance is regulated by the balance between stellar FUV and gas density. The CN column density then starts to decrease following the overall gas density profile. Consequently, the CN column density peak is reached at larger radii for higher values of $R_{\rm c}$. The same occurs for the radial intensity profile: the ring radius moves increasingly outward as $R_{\rm c}$ increases from 15 to 60 AU, and becomes almost constant for larger disk radii (Fig. \ref{fig:int_rcs} and \ref{fig:rings_rcs}). Since a high percentage of the intensity comes from the outer disk, the integrated fluxes increase with $R_{\rm c}$ (Fig. \ref{fig:fluxes_rcs}), and the smallest disk shows an order of magnitude lower intensity than the largest. It should be noted that there is a difference in mass between the $R_{\rm c}$=120 AU disk and for the 15 AU of about a factor of ten. However such a difference alone would only provide a variation of a few in flux (Fig. \ref{fig:ttau_flux_mass}).

\subsection{Varying the carbon abundance}
Recent observations of protoplanetary disks have shown that lines from CO and its isotopologues, the molecules usually adopted to measure the disk mass, are much weaker than expected in many disks \citep[e.g., ][]{favre2013,mcclure2016,miotello2016,pascucci2016,long2017,trapman2017}. This depletion of CO is inferred after photodissociation and freeze-out have been accounted for. One interpretation is that  CO is transformed to other less volatile molecules, which is mimicked in our models with an overall volatile C depletion. The gas-phase carbon abundance therefore has recently become one of the largest uncertainties in protoplanetary disks. To account for this effect, we ran three additional models of the $M_{\rm gas}=10^{-2}\,M_\odot$, $\psi=0.2$ and $h_{\rm c}=0.1$ T Tauri disk. In the first two, the initial overall abundances of carbon, oxygen, sulphur and nitrogen are depleted by a factor of 10 and 100. In the third model, carbon, nitrogen and sulphur are again depleted by a factor of 100, while oxygen is even more depleted in order to get $\rm [C]/[O]=1.5$. This assumption is motivated by \citet{kama2016} and \citet{bergin2016}, where a ratio $\rm [C]/[O]>1$ has been used to fit the C$_2$H flux in TW Hya. As shown in Fig. \ref{fig:fluxes_depletions}, the CN flux only varies by a factor of less than 2.5 in the four models in spite of two order of magnitude C and O abundance variations, thus showing a weak dependence of CN flux on the level of C and O abundance. 

This happens because as the CN abundance in the surface layers decreases, this abundance increases deeper in the disk, because of the lower amount of O, which is the species that CN is mainly destroyed by. For this reason, CN is not a good tracer of C and O depletion in protoplanetary disks.

\section{Discussion}\label{sec:discussion}

The integrated fluxes in our models are consistent with single-dish and interferometric observations. \citet{guilloteau2013} measured fluxes ranging from a few hundred $\rm mJy\,km\,s^{-1}$ up to $\sim10\rm\, Jy\,km\,s^{-1}$, which is exactly the range of values observed in the models. The same holds for the SMA interferometric observations in \citet{oberg2010,oberg2011}. \citet{oberg2011} stated that the radial profiles show different extents for CN and HCN in different disks. In particular, CN appears more radially extended than HCN in V4046 Sgr and LkCa 15. Using more recent ALMA observations, \citet{guzman2015} also concluded that CN is $\sim2$ times more extended than HCN in the disk surrounding the Herbig star MWC 480. This feature is well reproduced in our models (see Fig. \ref{fig:cnabu}). 

\begin{figure}[]
\centering
  \includegraphics[width=.5\textwidth]{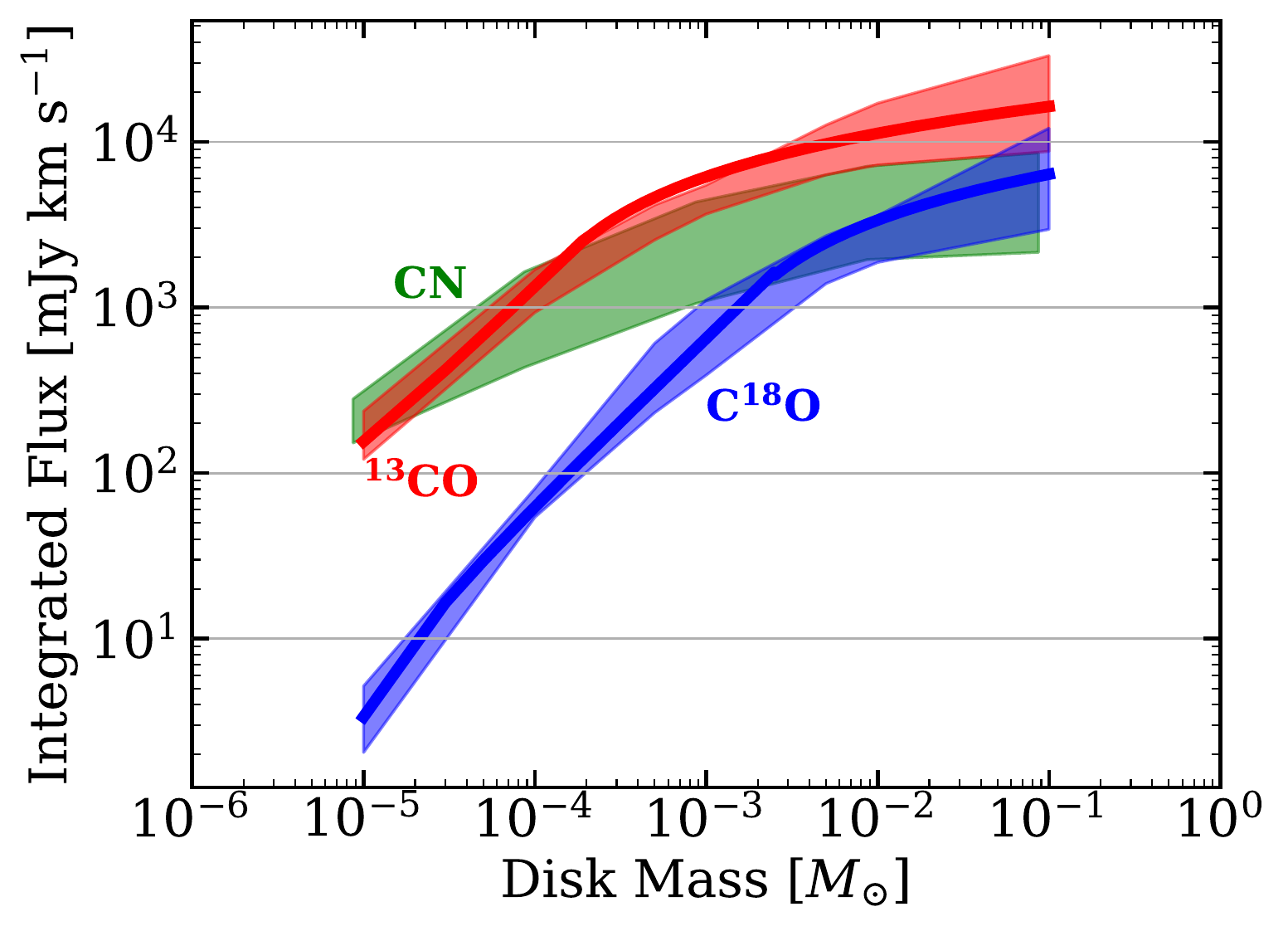}
\caption{Comparison between CN, $^{13}$CO and C$^{18}$O integrated flux dependence on the total disk mass. The $^{13}$CO and C$^{18}$O fluxes are taken from \citet{miotello2016}. The solid lines represent the fit functions of the median for $^{13}$CO (red) and C$^{18}$O (blue).}\label{fig:cnvsco}
\end{figure}

Finally, TW Hya is so far the only disk in which the \mbox{CN} emission has been observed and spatially resolved. CN shows a ring located close to 47 AU radius. With a mass of $0.056\,\rm M_\odot$ \citep{bergin2013} and a flaring index $\psi=0.3$ \citep{kama2016}, both the measured integrated flux for the CN ($N,J$)=($3-2,7/2-5/2$) transition of $27.1\,\rm Jy\, km\, s^{-1}$ \citep{kastner2014} and the the ring size is only slightly lower/smaller than the $\psi=0.3$ models we show in Fig. \ref{fig:ttau_flux_mass}. Still, those values are perfectly consistent with the fact that TW Hya is a low accretor, with $\dot M\sim10^{-9}\,\rm M_\odot\, year^{-1}$  \citep{debes2013}. For our models we assumed instead $\dot M=10^{-8}\,\rm M_\odot\, year^{-1}$, and, as shown in Fig. \ref{fig:fluxes_accretions}, smaller rings and lower fluxes are expected for lower accretion rates.

Recently, \citet{bergin2016} have shown that in TW Hya other  PDR tracers such as C$_2$H and c-C$_3$H$_2$ also present ring-like structures in their emission morphology. In order to reproduce such emission, \citet{bergin2016} concluded that a strong UV field combined with [C]/[O]>1 in the upper disk are needed. These conditions are not required for CN, and its ring shape is a natural consequence of the CN formation mechanism and therefore is expected to be a common feature observable in many full disks. The ring and brightness of the CN rings can however be enhanced by a higher UV flux impinging on the disk.

CN emission is therefore a good diagnostic for those parameters affecting UV flux. Among all, flaring is a particularly interesting case. The flaring index is usually derived by fitting the mid-IR range of the SED emitted by the inner few AU of the disk and this value is then assumed to be uniform out to the outer disk. For the first time, information about flaring at much larger radii can be gathered by studying the CN ring radius and flux when some constraints about the spectral type of the central star are available.

CN flux is however not a good tracer for the mass of the disk and CO isotopologues still provide a more sensitive diagnostic (see Fig. \ref{fig:cnvsco}). This is because in our models CN is most abundant in the upper layers of the disk, and no emission originates from the midplane where the bulk of the gas mass is. Our model results therefore do not support the results of \citet{chapillon2012}, who suggested the existence of large quantities of gas-phase CN close to the disk midplane, based on
the low measured values of the excitation temperature with pre-ALMA
data.

\section{Conclusions}

We modelled the CN abundance and emission in full protoplanetary disks around Herbig and T Tauri stars with the 2D thermochemical code DALI to self-consistently solve for the gas temperature structure and molecular abundance. The mass of the disks was varied between $10^{-5}$ and $10^{-1}\,\rm M_\odot$. We also varied other physical parameters, such as the flaring of the disk, and tested the effect of disk size, UV excess due to accretion onto the central star, and level of depletion of volatile C and O. The modelling shows the following results:

\begin{enumerate}
\item CN shows ring-shaped emission even in full protoplanetary disks. This is due to the formation route of CN, which is triggered by the excitation of H$_2$ to H$_2^*$ that is mostly abundant in a ring-shaped region close to the surface layer of the disk. H$_2$ is excited to  H$_2^*$ mostly through UV pumping. For this reason CN emission is very sensitive to UV flux.
\item Flaring plays an important role: disks with larger $\psi$ show systematically higher fluxes and larger rings. On the other hand, the UV intensity of the central stellar spectrum is critical itself. Herbig stars generally have stronger UV fields, and are thus expected to present larger disks and higher integrated fluxes than T Tauri stars. As for T Tauri stars, UV excess can in first approximation  be quantified through the accretion rate \citep[e.g., ][]{kama2016}, with higher accretion rates producing higher UV excess fluxes and consequently brighter and larger CN rings.
\item The integrated fluxes of the models increase as the total disk mass increases but CN is still less sensitive than the CO isotopologues to the total gas mass. On the other hand, ring size is a much easier parameter to measure and, especially for Herbig stars, it is strongly dependent on the total gas disk mass, therefore providing a useful additional parameter to put constraints on the mass of disks around Herbig stars.  
\item The disk critical radius $R_{\rm c}$ only plays a role in the CN morphology when smaller than 60 AU: in this case, smaller disks also show smaller CN rings and lower integrated fluxes. However, disks with $R_{\rm c}$ larger than 60 AU do not show any qualitative difference from disks with $R_{\rm c}=60\rm\, AU$. The disk scale height does not strongly affect the CN emission.
\end{enumerate}

The dependence of CN flux and ring size on the above disk parameters make it a sensitive probe of the disk structure. In particular, when the spectral type of the central star is known CN can be combined with the SED to obtain information about parameters that are usually poorly constrained, such as the flaring of the disk at large radii, and to get an independent estimate of the  UV flux.

\begin{acknowledgements}
We thank the anonymous referee for his/her constructive comments. We also thank A. Miotello, S. van Terwisga, C. Walsh, A. Sternberg, A. Faure, F. Lique andN. van der Marel for useful discussions. Astrochemistry in
Leiden is supported by the European Union A-ERC grant 291141 CHEMPLAN,
by the Netherlands Research School for Astronomy (NOVA), and by a Royal
Netherlands Academy of Arts and Sciences (KNAW) professor prize. All the figures
were generated with the \texttt{python}-based package \texttt{matplotlib} \citep{hunter2007}.
\end{acknowledgements}

\bibliographystyle{aa}

\begin{appendix}

\section{Constant $G_0/n_{\rm gas}$}
\begin{figure}[h!]
\centering
  \includegraphics[width=\linewidth]{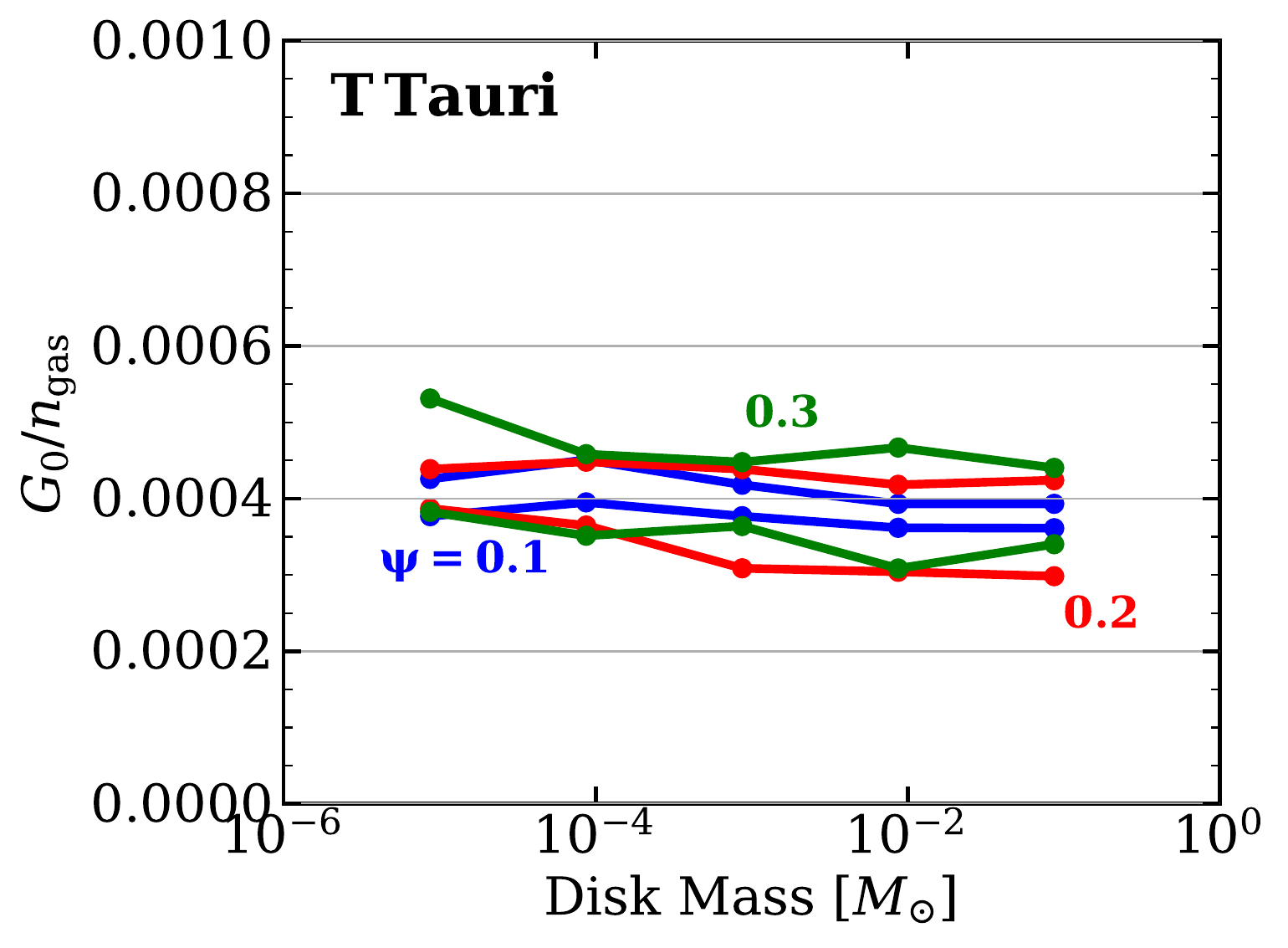}
\caption{Value for $G_0/n_{\rm gas}$ where the CN abundance peaks for all the T Tauri models. The different lines for each colour refer to models with different $h_{\rm c}$ . }\label{fig:g0ngas}
\end{figure}

\begin{figure}[h!]
\centering
\includegraphics[page=1,width=\linewidth]{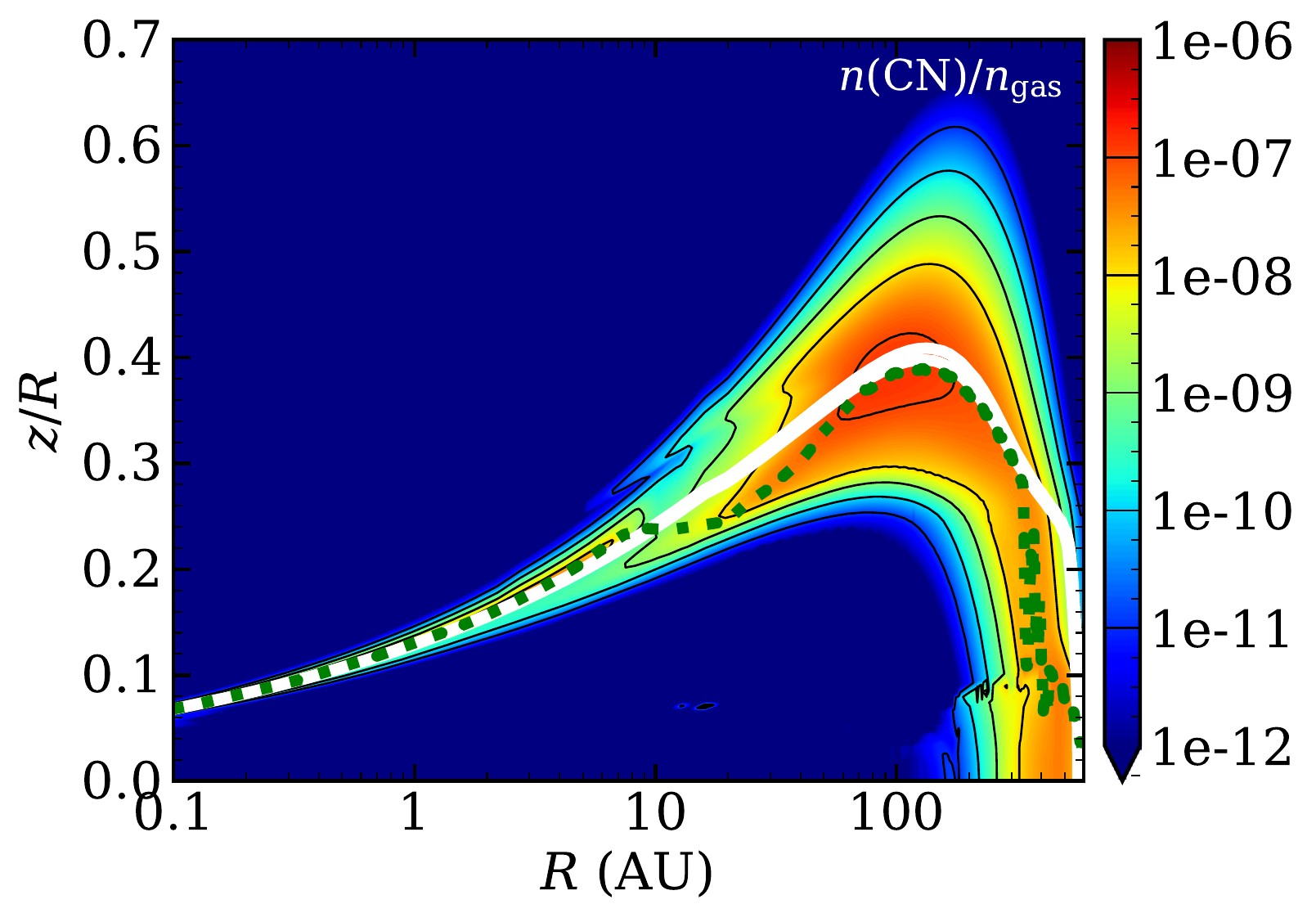}
\caption{Overlay of the CN abundance structure in a  $10^{-2}\,\rm M_{\odot}$, $\psi=0.3$, $h_{\rm c}=0.1$ disk surrounding a T Tauri star with a green dotted line indicating the location of the vertical CN abundance maximum for each radius. The white line highlights a surface of constant $G_0/n_{\rm gas}$. The two lines follow each other very closely.}\label{fig:g0ngasbis}
\end{figure}

In all of our models, the value of G$_0/n_{\rm gas}$ is constant to within a factor of a few. Fig. \ref{fig:g0ngas} shows this value for our T Tauri models and the same holds for Herbigs.  In Fig. \ref{fig:g0ngasbis} not only the abundance peak, but at each radius the CN abundance maximum (dotted green line) follows the constant G$_0/n_{\rm gas}$ profile (white line). The model represents a  $10^{-2}\,\rm M_{\odot}$, $\psi=0.3$, $h_{\rm c}=0.1$ around a T Tauri star.

\section{Density profiles for different values of $R_{\rm c}$}

\begin{figure}[h!]
\centering
  \includegraphics[width=\linewidth]{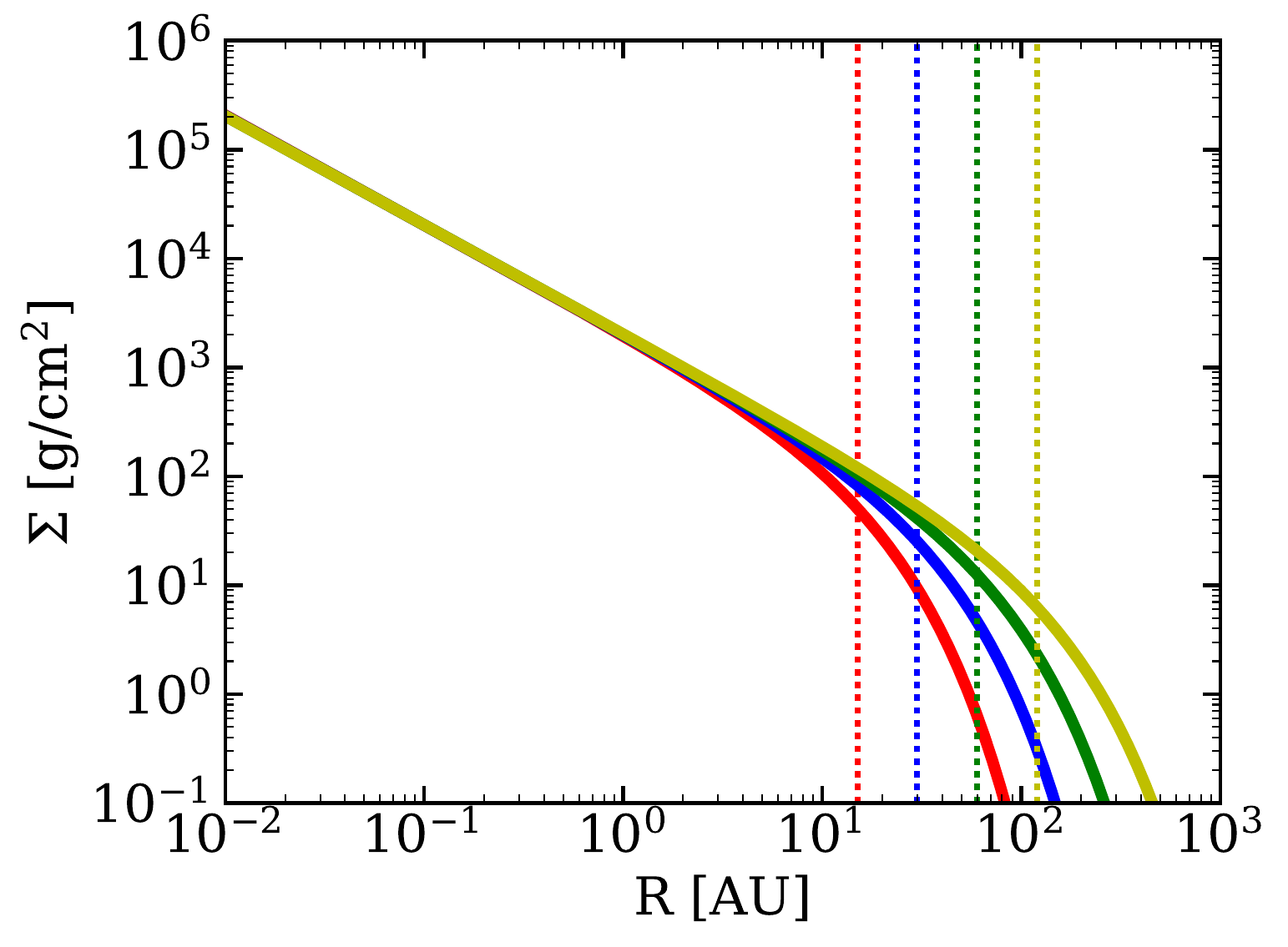}
\caption{Comparison between the radial surface density profiles used when comparing models with different $R_{\rm c}$ parameters. The dotted line indicates, for each profile, the location of $R_{\rm c}$.}\label{fig:denscomp}
\end{figure}

When testing the effect of the disk size on CN emission, we varied the values of $R_{c}$ keeping the density profiles at the small radii equal for all the models, as shown in Fig. \ref{fig:denscomp}. This keeps the conditions in the inner disk equal for all the models.

\section{Level excitation and ring location}\label{app:levelpop}

As shown in Fig. \ref{fig:columndens}, and as observed also in other disk models around Herbig stars, the CN (N=3-2) emission peaks at radii which are larger than that of the column density peak. This is an excitation effect that originates from the high critical densities of the CN transitions, combined with the low energies of the transitions.

\begin{figure}[h!]
\centering
  \includegraphics[width=\linewidth]{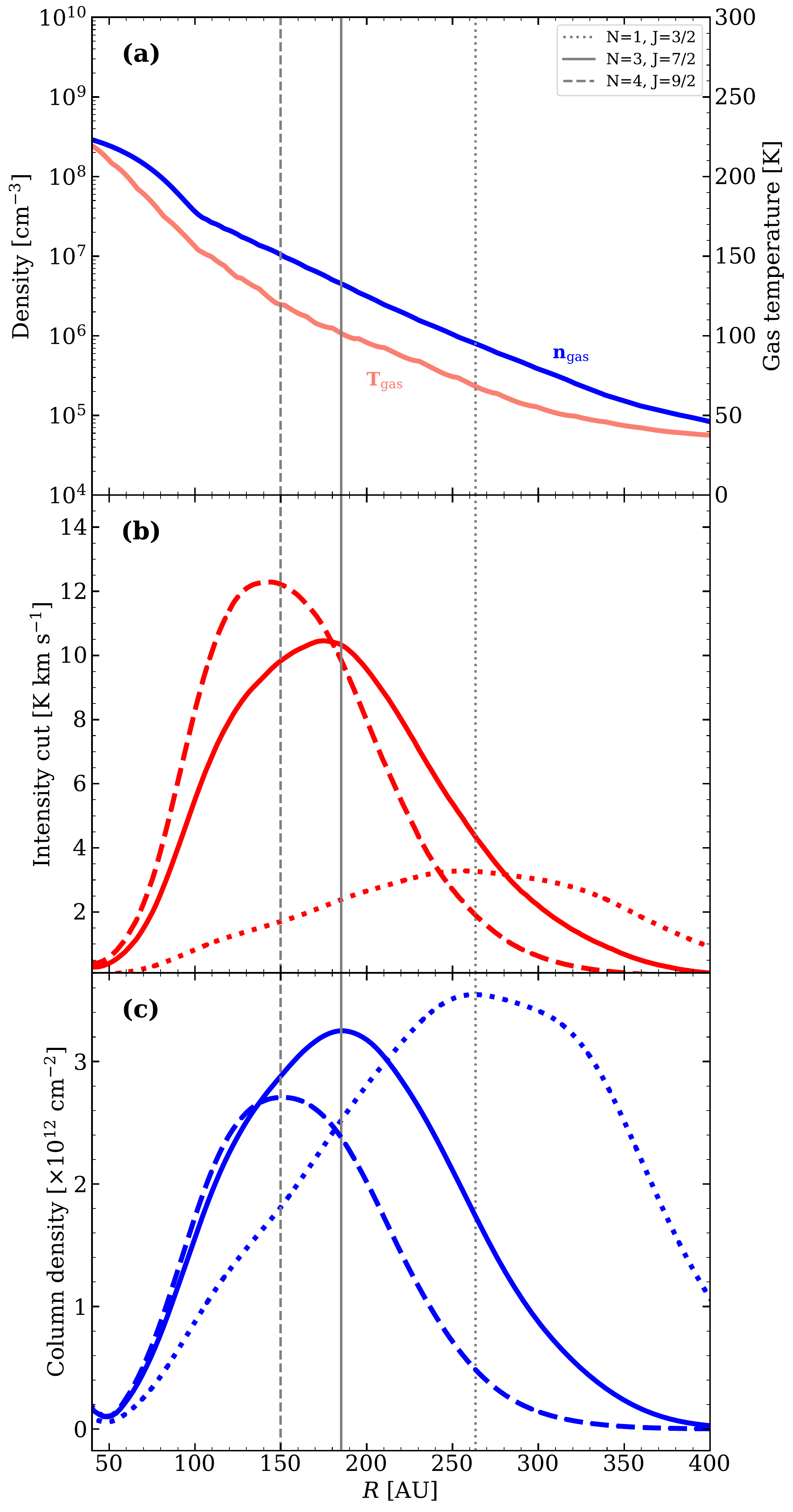}
\caption{(a) Gas density and temperature along the CN high abundance layer (dotted-green line in Fig. \ref{fig:cnabu}). (c) Column densities of CN molecules in the (N=1, J=3/2) (dotted), (N=3, J=7/2) (solid) and (N=5, J=11/2) (dashed) levels. (b) Intensities of the N=4-3, N=3-2 and N=1-0 transitions, for comparison with panel (c). The vertical lines mark the location of the column density peaks of CN molecules in the three levels.}\label{fig:lev_pops}
\end{figure}

Two main effects play a role:
\begin{enumerate}
\item The CN column density decreases more slowly toward the outer radii for the Herbig disks than for the T Tauri ones (see Fig. \ref{fig:columndens}).
\item The gas density along the CN high-abundance layer decreases with radius (Fig. \ref{fig:lev_pops}a). Gas temperature also decreases, but  remains always above the energy of the N=3-2 transitions ($\sim$ 30K) and also of the N=4-3 one ($\sim$ 50K).
\end{enumerate}

What happens is then the following:

\begin{itemize}
\item In the inner region of the disk, the gas density and temperature are high enough for the lines to be thermalized: as the column density increases, the intensity increases accordingly.
\item Moving outward along the CN layer, even though the gas temperature remains high, the gas density decreases: as the peak of the CN column density is reached (\mbox{$\sim 120\,\rm AU$} in this models) the gas density has fallen below the critical density of the higher level transitions (\mbox{$n_{\rm crit} = 9.12\times 10^6\,\rm cm^{-3}$} at 50 K for the N=4-3, J=9/2-7/2 transition). Consequently, higher levels get depopulated and the population of the lower levels grows at larger radii t and at lower gas densities (see Fig. \ref{fig:lev_pops}c). 
\item Since CN is optically thin, the emission intensity only depends on the population of the upper layers of the transition. The emission thus follows the increase/decrease of populations (Fig. \ref{fig:lev_pops}b).
\item Moving even further out, at some radius the critical density of the N=3-2 transition (\mbox{$n_{\rm crit} = 3.84\times 10^6\,\rm cm^{-3}$} at 50 K for the N=3-2, J=7/2-5/2 transition) is also reached: as explained above level 3 gets unpopulated, the column density of CN in level N=3 decreases, 3-2 intensity starts decreasing and the population of levels 2 and 1 increases.
\end{itemize}

This effect is not evident for the T Tauri stars: as mentioned above, for their disks the CN column density profile decreases more steeply, and dominates over the increase in population of the lower levels.

\end{appendix}

\end{document}